\documentclass[12pt,a4paper,final]{iopart}
\usepackage{cite}
\usepackage{color}
\usepackage{bm}
\usepackage{float}
\usepackage{stmaryrd,scalerel} 
\usepackage{iopams}
\usepackage{color}
\usepackage[breaklinks=true,colorlinks=true,linkcolor=blue,urlcolor=blue,citecolor=blue]{hyperref}
\expandafter\let\csname equation*\endcsname=\relax
\expandafter\let\csname endequation*\endcsname=\relax
\usepackage{amsmath}

\newenvironment{claim}[1]{\par\noindent\underline{Claim:}\space#1}{}

\begin{document}

\title{Stochastic gyration driven by dichotomous noises}
\author{Timoth\'ee Herbeau$^{1}$, Leonid  Pastur$^{2,3}$, Pascal Viot$^{1}$, and Gleb Oshanin$^{1}$}
\address{$^1$ Sorbonne Universit\'e, CNRS, Laboratoire de Physique Th\'eorique de la Mati\`ere Condens\'ee (UMR CNRS 7600), 4 Place Jussieu, 75252 Paris Cedex 05, France}
\address{$^2$  King's College London, Department of Mathematics, Strand, London WC2R 2LS, UK}
\address{$^3$  B. Verkin Institute for Low Temperature Physics and Engineering, $47$
Nauky Ave., Kharkiv, 61103, Ukraine}

\begin{abstract}
	We consider stochastic dynamics of a particle on a plane in presence of two noises and 
	a confining parabolic potential - an analog of the experimentally-relevant Brownian Gyrator (BG) model. In contrast to the standard BG model, we suppose here that the time-evolution of the position components is driven not by Gaussian white-noises, but by two statistically-independent dichotomous noises. We calculate analytically the position variances and cross-correlations, as well as the mean angular momentum, which permits us to establish the conditions in which a spontaneous rotational motion of the particle around the origin takes place. We also present a numerical analysis of the mean angular velocity.  
	Lastly, we 
	calculate analytically some marginal position probability density functions revealing a remarkably rich behavior that emerges in such a system of two coupled linear stochastic differential equations.  We show that depending on the values of  parameters characterizing noises these distributions approach the steady-state forms defined on a finite support, having very unusual shapes, possessing multiple maxima and minima, plateaus and exhibiting a discontinuous behavior.  
\end{abstract}

\vspace{0.2in}
Keywords: Out-of-equilibrium dynamics, the generalized  Ornstein-Uhlenbeck processes, dichotomous noises, stochastic gyration

\vspace{0.2in}


\maketitle

\section{Introduction}

Out-of-equilibrium systems exhibit rich, often counter-intuitive dynamics that lie beyond the scope of traditional equilibrium statistical mechanics. These systems can sustain currents, perform mechanical work, and  maintain steady states far from thermodynamic equilibrium. Comprehensive reviews linking stochastic dynamics, transport, and fluctuations 
can be found in \cite{Seifert2012,Peliti2021,Mejia2025,Sarracino2025}. Notable examples of out-of-equilibrium systems include biological molecular motors, active cytoskeletal transport, and synthetic nanomachines, where stochastic fluctuations, energy dissipation, and mechanochemical couplings play critical roles \cite{Reimann2002,Astumian2002,Berg2003,Hanggi2009,Sekimoto2010,Bechinger2016,Ciliberto2017,Li2019,Sou2019}. The state of art in the design of synthetic out-of-equilibrium molecular systems, from dissipative supramolecular assemblies to light-driven rotary motors was recently reviewed in \cite{Giuseppone2021}.

A quite simple  out-of-equilibrium system which exhibits 
some of the aforementioned features is the so-called Brownian Gyrator (BG) model. The model consists of a particle moving randomly on a plane in presence of a parabolic potential and being subject to two statistically-independent Gaussian white-noises with unequal amplitudes (temperatures), acting along the $x$- and $y$-directions, respectively. Mathematically, the BG model is framed in terms of two linearly coupled Ornstein-Uhlenbeck processes, each evolving subject to its own white-noise. First introduced in \cite{Exartier1999} for the analysis of effective temperatures in out-of-equilibrium system, 
this model was revisited in \cite{Filliger2007}, where it was claimed that if the temperatures of the two heat bathes are not equal, the particle experiences  
a \textit{systematic} torque. To support this claim, it was shown in \cite{Filliger2007} that indeed the \textit{mean} torque is non-zero under such out-of-equilibrium conditions and therefore, the BG can be deemed as a minimal stochastic model of a microscopic heat engine that performs, on average, a rotational motion. The BG model has been experimentally realized  
by constructing equivalent electrical circuits \cite{Ciliberto2013,Ciliberto2013a} (see also \cite{Cerasoli2022a}) or by immersing 
an optically-trapped colloidal particle in a bath with different temperatures acting along the perpendicular directions and maintaining therefore a non-equilibrium steady-state \cite{Argun2017}, which  permitted to  confirm some  
of theoretical predictions.

Due to a non-trivial dynamical behavior emerging in a very simple setting, the BG model has attracted a great deal of attention. 
Subsequent analyses 
determined the mean specific angular velocity \cite{Dotsenko2013,Mancois2018,Bae2021} and scrutinized 
various facets of the dynamical and steady-state behaviors for the BG model in the quantum regime \cite{Fogedby2018}, 
for delta-correlated in time noises \cite{Ciliberto2013,Ciliberto2013a,Crisanti2012,Grosberg2015,Lahiri2017, Cerasoli2018,Dotsenko2019,Tyagi2020,Chang2021,Siches2022,Lucente2022,Abdoli2022},
 and also for the Gaussian noises with long-ranged temporal correlations \cite{Nascimento2021,Squarcini2022a}. 
Apart of that, effects of inertia 
on the performance of the BG \cite{Bae2021} (see also \cite{Dotsenko2024}) and of 
an anisotropy of fluctuations \cite{MovillaMiangolarra2021,Miangolarra2022} have been studied. Moreover, it was realized that the probability currents in the steady-state are non-vanishing and moreover, possess non-zero curls such that
 in addition to the rotational motion around the origin, the BG is ongoingly spinning around its center of mass \cite{Cerasoli2018}. Quite remarkably, any two BG particles coupled by hydrodynamic interactions spin in a completely synchronized way \cite{Dotsenko2022a,Dotsenko2023}. Lastly,     
the BG model was used to test some basic concepts of stochastic thermodynamics \cite{Berut2014,Berut2016,Ciliberto2013,Ciliberto2013a,Cerasoli2018,Mazzolo2023}.

Concurrently, it is evident that since the BG  
operates in heat baths, the 
torque, the angular momentum and the angular velocity are random variables which 
fluctuate in time and also vary randomly from one realization of the process to another. Up to a recent time, only the \textit{mean} values of these random variables were known, although it is clear that such a knowledge is insufficient to fully characterize the dynamics. This issue has been addressed in recent \cite{Viot2024}, in which the full probability density functions (PDFs) of these random variables have been determined for a time-discretized stochastic dynamics. In particular, it was shown analytically  and confirmed experimentally that the PDF $P(W)$ of the angular velocity $W$ has heavy power-law tails $1/|W|^3$ meaning that, in fact, the first moment of $W$ is the \textit{only} existing moment. In other words, fluctuations destroy completely any systematic rotational motion and the BG cannot be viewed as a common-sense heat engine. Note that exactly the same large-$W$ behavior has been obtained in \cite{Dotsenko2024} for the PDF of $W$ for a two-dimensional stochastic acceleration process in the same 
parabolic potential, which may suggest that it is a generic feature of the BG-like processes driven by Gaussian white-noises. 
In \cite{Viot2024}, it was argued that such heavy tails emerge 
due to the fact that for the BG driven by Gaussian noise the BG position PDF always has a maximum at the origin, where the moment of inertia is zero and hence, the angular velocity becomes infinitely large.
Consequently, a certain "tuning"  of the standard BG model is desirable in order to get a stochastic motor with a more systematic performance.

In the present paper we revisit the standard BG model assuming that in place of the Gaussian white-noises acting on the components, one has symmetric \textit{dichotomous} noises. Dynamics of diverse physical systems subject to such noises has been widely studied (see, e.g., \cite{Fitzhugh1983,Kampen2007,Lifshits1988,Haeunggi1994,Zhou2010,Klyatskin2011,Benabdallah2022} and references therein) and also re-gained much interest recently, especially within the context of the so-called run-and-tumble dynamics observed in active systems for which many remarkable results have been obtained (see, e.g., \cite{Zakine2017,Gradenigo2019,Dean2021,Santra2023,Trajanovski2023a,Loewe2024,Gueneau2025,Santra2021,Sandev2022,Smith2022,Dutta2024,Basu2024,Frydel2025}). Here, our focal interest  
is to specify a possible range of parameters of dichotomous noises, if any, 
in which the particle's position PDF will have a minimum (or desirably, a zero value) at the origin, such that one may expect to encounter a more regular behavior of the angular velocity and hence, a more systematic rotational motion.   
From a bit different perspective, the model to be studied here is quite interesting due to the following reason:  
 in the standard BG model each of the position components is driven by a Gaussian white-noise - an equilibrium "thermal" noise, such that in the absence of coupling between the two components each of them is in detailed balance with its own heat bath maintained at some fixed temperature and satisfies the fluctuation-dissipation theorem. For the non-zero coupling between the two components, the entire system evolves towards an out-of-equilibrium steady-state with an emerging  rotational motion solely in case when the temperatures (noise rates) of the respective heat baths  are not equal to each other, such that there is no unique temperature in the system. Once the temperatures of two heat baths are equal, the system reaches an equilibrium and no rotational motion takes place. In contrast, dichotomous noise inherently acts as an out-of-equilibrium driving mechanism and, in general, entails non-zero internal currents violating the detailed balance.  
 
 We note finally that our model can be apparently realized experimentally. As a matter of fact, 
 the noises imposed on the optically-trapped particle \cite{Argun2017}  or on the resistances in the electric circuit \cite{Ciliberto2013,Ciliberto2013a} were generated by a computer, such that, in principle, noises with any given 
statistical properties  can be designed.

In this paper we concentrate only on some particular aspects of the dynamical behavior: we calculate the moments and the cross-moment of the position components, the mean values of the angular momentum (analytically) and of the angular velocity (numerically), determining the conditions 
under which the mean angular momentum and the mean angular velocity 
are not equal to zero:  
hence, a stochastic rotational motion around the origin may, in principle, take place.
We focus next on the marginal position PDF in a coordinate system rotated on the angle $-\pi/4$, which appears to be a mathematically less involved problem than the derivation of the marginal and of the full PDFs. However, this proves to be sufficient to gain an insight into the dynamical behavior and to show that in case of dichotomous noises the behavior becomes by far richer than in the standard BG model with Gaussian white-noises. We proceed to demonstrate that, depending on the parameters characterizing the noises, the marginal position PDFs converge to steady-state forms supported on a finite domain and having quite unusual shapes with multiple extrema and discontinuities. 
Notably, in some regimes the probability of finding the particle at the origin in the steady-state is exactly zero suggesting that a heat engine with a more systematic performance can be indeed realized.  
A more detailed analysis with 
exact results on the full position PDF and the PDFs of the angular momentum and of the angular velocity
 will be presented elsewhere \cite{Herbeau2026}.

The paper is organized as follows: In Sec. \ref{mod} we formulate our model, introduce basic notations and the random variables under study. In Sec. \ref{mom} we calculate the second moments of positions along the $x$ and $y$ axes and their cross-moment. 
Further on, we determine the first moment of the specific angular momentum, and discuss the results of a numerical analysis of the mean specific angular velocity. 
 In Sec. \ref{PDF} we present some general results on the limiting ($t = \infty$) form of the PDFs $P(\mathcal{S } = x + y)$  and $P(\mathcal{D} = x - y)$, which can be thought of as the marginal PDFs in the coordinate system $(x',y')$ obtained from the original one by rotating it on the angle $-\pi/4$. Here we also specify six different regimes in which the PDFs may have different shapes, multiple extrema and exhibit a discontinuous behavior. 
Next, in Sec. \ref{part} we give exact results for the PDFs
 for each of these six regimes by choosing some
representative values of the noise switching rates for which the Fourier transform can be obtained in a closed form. 
Further on, we consider in Sec. \ref{Zx0}  the situation in which either of the noises is absent.  
Finally, we conclude in Sec. \ref{conc} with a brief summary of our findings.  For completeness, 
in \ref{A} we present the derivation of the position probability density function of the generalized Ornstein-Uhlenbeck process driven by a dichotomous noise. In \ref{GG} we discuss  
the \textit{mixed} case when either of the noises is a Gaussian white-noise, while the other one is a dichotomous noise.

\section{Model and basic notations}
\label{mod}

Consider a particle that moves on the $(x,y)$-plane in presence of a parabolic 
confining potential $U$ and is subject to random forces  $\zeta_x(t)$ and $\zeta_y(t)$ that act independently along the $x$- and $y$-axes.  

The overdamped dynamics of the particle is described by two coupled stochastic differential  equations 
\begin{align}
\label{LE}
\begin{split}
\eta \dot{x}(t) &= - \frac{\partial U}{\partial x} + \zeta_x(t) =  - x(t) + u \, y(t) + \zeta_x(t) \,, \quad x(0) = 0 \,,  \\
\eta \dot{y}(t) &= - \frac{\partial U}{\partial y} + \zeta_y(t) = - y(t) + u \, x(t) + \zeta_y(t) \,, \quad y(0) = 0 \,,
\end{split}
\end{align}
where  the dot denotes the time-derivative and the confining potential is 
\begin{align}
U = \frac{x^2}{2} + \frac{y^2}{2} - u x y \,, \quad |u| <  1 \,.
\end{align}
Note that once $u=0$, the equations \eqref{LE} decouple and describe two independent stochastic (generalized Ornstein-Uhlenbeck \cite{Caceres1997}) processes driven by random forces $\zeta_x(t) $ and $ \zeta_y(t)$, respectively. For $|u| < 1$ the particle is localized, while for $|u| \geq 1$ it gets delocalized and travels to infinity. When $\zeta_x(t)$ and $\zeta_y(t)$ are independent Gaussian white-noises, the above model is the standard BG model framed in terms of two coupled standard Ornstein-Uhlenbeck processes. In what follows, we will call the model in eqs. \eqref{LE} the Stochastic Gyrator (SG).

The forces 
$\zeta_x(t)$ and $\zeta_y(t)$ are statistically-independent symmetric \textit{dichotomous} noises (see, e.g., \cite{Kampen2007,Lifshits1988,Haeunggi1994,Benabdallah2022}) that switch randomly between two values $\pm v_{a}$, where here and henceforth $a = x,y$,
\begin{align}
\label{noise1}
\zeta_{a}(t) = \begin{cases}
+ v_{a}, \quad \text{prob = $1/2$},\\
- v_{a},  \quad \text{prob = $1/2$}.
\end{cases}
\end{align}
Each of $\zeta_{a}(t)$ is a Markov process, hence, 
the transitions between their two states are memoryless and the time-intervals between successive switching events
are independent, exponentially distributed random variables. The respective switching rates are denoted by $\lambda_a$. 
Consequently, the mean and the autocorrelation of the noises obey 
\begin{align}
\begin{split}
\label{noise2}
&\mathbb{E}_{a}\left[ \zeta_{a}(t) \right] = 0 \,, \\
&\mathbb{E}_{a}\left[ \zeta_{a}(t) \zeta_{b}(t')\right] = \delta_{a,b} v^2_{a} \exp\left( -2 \lambda_{a} |t - t'| \right) \,,
\end{split}
\end{align}
where the symbol $\mathbb{E}_{a}\left[ \ldots \right]$ denotes averaging with respect to realizations of the corresponding 
dichotomous noise, while $\delta_{a,b}$, $a,b = x,y$, is the Kronecker symbol such that $\delta_{a,b} = 1$ for $a=b$ and zero, otherwise. Note that the autocorrelation function in the second line in eq. \eqref{noise2} is also customarily 
written as
\begin{align}
	\label{diff}
\mathbb{E}_{a}\left[ \zeta_{a}(t) \zeta_{b}(t')\right] = \delta_{a,b}\frac{D_{a}}{\tau_{a}} \exp\left( - \frac{ |t - t'|}{\tau_{a}} \right) \,,
\end{align}
with $\tau_{a} = 1/ (2\lambda_{a})$ being the finite correlation time of the noise processes, while the amplitudes $D_{a}$, $a=x,y$, are the "diffusion coefficients", i.e., the proportionality factors in the linear dependencies on time of the corresponding mean-square displacements of the particle along the $x$ and $y$ axes, respectively, in absence of the potential U, 
\begin{align}
\label{diffD}
D_{a} = \frac{v_{a}^2}{2 \lambda_{a}} \,.
\end{align}
It is worth noting that in virtue of eq. \eqref{diff} the dichotomous noise converges to Gaussian white-noise in the limit 
$\tau_a \to 0$ with 
$D_a$
held constant, i.e., by taking 
$v_a \to \infty$ 
and 
$\lambda_a \to \infty$
 simultaneously, while keeping their ratio in eq. \eqref{diffD} fixed. We will use this limit to benchmark our results for dichotomous-noise dynamics against those of the standard BG model.

\begin{figure}[H]
	\begin{center}
\includegraphics[width=5cm]{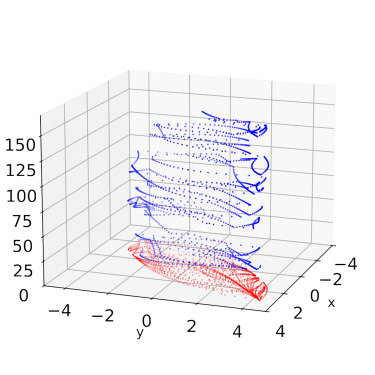}
\includegraphics[width=5cm]{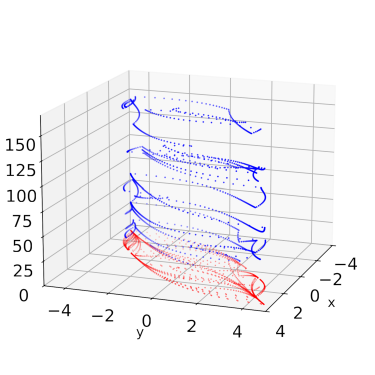}
\includegraphics[width=5cm]{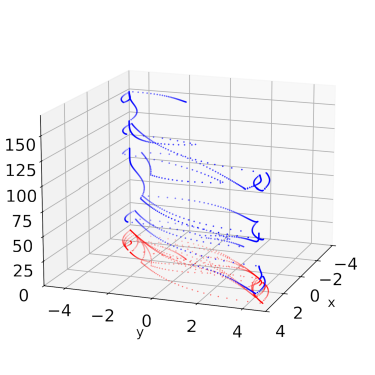}
	\end{center}
	\caption{Three randomly-generated SG trajectories (blue curves) shown in the 
		$(x,y,t)$
		space for 
		$v_x=1$ and
		$v_y=3$, with switching rates 
		$\lambda_x = \lambda_y = 1$ (left panel), 
		$1/3$
		(middle panel), and 
		$0.2$
		(right panel). The red curves represent the projections of these trajectories onto the 
		$xy$-plane. } \label{fig:000}
\end{figure}

The system of two coupled eqs. \eqref{LE} can be solved by standard means for the trajectories, for given realizations of noises, to give
\begin{align}
\label{a}
\begin{split}
x(t) & = \int^t_0 d\tau \, Q_c(t-\tau) \, \zeta_x(\tau) + \int^t_0 d\tau \, Q_s(t-\tau) \, \zeta_y(\tau) \,, \\
y(t) & = \int^t_0 d\tau \, Q_s(t-\tau) \, \zeta_x(\tau) + \int^t_0 d\tau \, Q_c(t-\tau) \, \zeta_y(\tau) \,,
\end{split}
\end{align}
where the kernels 
\begin{align}
\begin{split}
Q_c(t) = e^{-t} \cosh(u t) \,, \qquad 
Q_s(t) = e^{-t} \sinh(u t) \,.
\end{split}
\end{align}
It is important to note that, in contrast to the position components in the standard BG model, that are unbounded, more precisely, vary over the entire real line, the components 
$x(t)$
and 
$y(t)$
of the SG driven by dichotomous noises evolve within a bounded region. The size of this region gradually increases with time and eventually approaches a finite limit as 
$t \to \infty$. Indeed, replacing in  eqs. \eqref{a}
the noises $\zeta_a(t)$, $a = x,y$, by their 
amplitudes we obtain
\begin{align}
\label{bounded}
|x(t)| \leq \frac{v_x + u v_y}{1 - u^2} + {\rm const.} \, e^{-(1 - |u|) t} \,, \quad |y(t)| \leq \frac{u v_x + v_y}{1 - u^2} + {\rm const.} \, e^{-(1 - |u|) t} \,.
\end{align} 
Figure \ref{fig:000} shows a parametric plot in the 
$(x,y,t)$ space of three randomly generated trajectories, explicitly illustrating that the SG remains confined within a bounded “tube” along the 
$t$-axis.

Below we will calculate exactly the first two moments of these random variables and also their cross-moment $\mathbb{E}_{x,y}\left[x(t) y(t) \right]$, which will permit us to determine the first moment of the specific (per unit mass) angular momentum of the SG:
\begin{align}
\label{L}
L = x(t) \dot{y}(t) - \dot{x}(t) y(t) \,.
\end{align}
By analyzing the behavior of the mean specific angular momentum, we will identify the conditions under which it differs from zero -- indicating thereby that the SG exhibits a tendency to perform a stochastic rotational motion around the origin, akin to that observed in the BG.
We will also investigate the behavior of the SG’s specific angular velocity:
\begin{align}
\label{W}
W = \frac{x(t) \dot{y}(t) - \dot{x}(t) y(t)}{x^2(t) + y^2(t)}  \,.
\end{align}
This analysis will be carried out numerically - we will determine the first moment of $W$ and study its dependence on the parameters of noises.

Lastly, we will focus on some stationary  marginal  
position PDFs of the SG. Here, instead of focusing on the marginal PDFs $P(x)$ and $P(y)$ and on the full PDF $P(x,y)$, which will be discussed elsewhere \cite{Herbeau2026},
 we will address a somewhat simpler problem and calculate the PDFs of random variables
 \begin{align}
 \label{SD}
 \mathcal{S} & = x(t) + y(t) \,, \\
  \mathcal{D} & = x(t) - y(t) \,,
 \end{align}
 where $\mathcal{S}$ defines (up to the factor $1/2$) the position of the center-of-mass of two components, while $\mathcal{D}$ is their instantaneous relative position. Note that the probability density functions of  $\mathcal{S}$ and  $\mathcal{D}$
 can be also thought of as the marginal distributions in the coordinate system $(x',y')$ obtained upon a rotation of the original $(x,y)$ system on the angle $= - \pi/4$.  Note, as well, that likewise the position component $x(t)$ and $y(t)$ (see eq. \eqref{bounded}), the random variables $\mathcal{S}$ and $\mathcal{D}$ also have a finite support. We will discuss this point in more detail below.
  
In virtue of eqs. \eqref{LE}, the variables $\mathcal{S}$ and $\mathcal{D}$ obey
\begin{align}
\label{UV}
\begin{split}
\dot{\mathcal{S}}(t) &= - (1-u) \mathcal{S}(t) + \left(\zeta_x(t) + \zeta_y(t)\right) \,, \\
\dot{\mathcal{D}}(t) &= - (1+u) \mathcal{D}(t) + \left(\zeta_x(t) -   \zeta_y(t)\right) \,,
\end{split}
\end{align}
hence, their instantaneous values are 
\begin{align}
\begin{split}
\label{inst}
\mathcal{S}(t) &= \int^t_0 d\tau e^{-(1-u)(t - \tau)} \zeta_x(\tau) + \int^t_0 d\tau e^{-(1-u)(t - \tau)} \zeta_y(\tau) = \mathcal{S}_x + \mathcal{S}_y \,,\\
\mathcal{D}(t) &= \int^t_0 d\tau e^{-(1+u)(t - \tau)} \zeta_x(\tau) - \int^t_0 d\tau e^{-(1+u)(t - \tau)} \zeta_y(\tau) = \mathcal{D}_x - \mathcal{D}_y \,.\\
\end{split}
\end{align}
The kernels in the above integrals are exponential functions, as compared to eqs. \eqref{a}, which contain hyperbolic functions.  Thus, the  derivation of the corresponding PDFs is a mathematically much easier task than the derivation of the position PDFs. Concurrently, it will allow us to demonstrate that the behavior of the SG driven by dichotomous noises is by far richer than the behavior in the system with Gaussian white-noises, i.e., for the standard definition of the BG model. 
 
 \section{Moments of positions of the SG, specific angular momentum and specific angular velocity}
\label{mom}

We focus on the moments of $x(t)$ and $y(t)$, their cross-moment, and also of the specific angular momentum and the specific angular velocity.

\subsection{Moments of position components}

From Eqs. \eqref{a} one finds straightforwardly the second moments of $x(t)$ and $y(y)$ (the first moments evidently vanish)
\begin{equation}
	\begin{split}
		\label{momentsxy}
		\mathbb{E}_{x,y}\left[ x^2(t) \right] &= 
		 D_x  I_c(\lambda = \lambda_x)  + D_y I_s(\lambda = \lambda_y)\,,\\
		\mathbb{E}_{x,y}\left[ y^2(t) \right] 
		&= D_x I_s(\lambda = \lambda_x)  + D_y I_c(\lambda = \lambda_y) \,,\\
		\mathbb{E}_{x,y}\left[ x(t)y(t) \right] 
		&= D_x I_{sc}(\lambda = \lambda_x)  + D_y I_{sc}(\lambda = \lambda_y) \,,	
	\end{split}
\end{equation}
where $D_a$ are defined in eqs. \eqref{diffD}  and the functions $I_c(\lambda)$, $I_s(\lambda)$ and $I{sc}(\lambda)$ can be readily calculated explicitly for arbitrary $t$. In the limit $t \to \infty$ these functions attain simple limiting forms:
\begin{align}
	\label{I}
	\begin{split}
		I_c(\lambda) &= 2\frac{(1 + (2 - u^2) \lambda)\lambda}{(1-u^2) ((1 + 2\lambda)^2 - u^2)}  \,, \quad \\
		I_s(\lambda) &= 2\frac{(1+ \lambda) \lambda u^2} {(1-u^2) ((1 + 2\lambda)^2 - u^2)} \,,\\
		I_{sc}(\lambda) &= 2\frac{u\lambda (1+\lambda)}{(1-u^2) ((1 + 2\lambda)^2 - u^2)}\, 
	\end{split}
\end{align}

In the Gaussian white-noise limit, (see the text below eq. \eqref{diffD}), one recovers from the analogs of 
 eqs. \eqref{I} the expression for the standard BG model \cite{Dotsenko2013}, in which  
\begin{equation}\label{eq:aux3}
			\begin{split}
				\mathbb{E}_{x,y}\left[ x^2(t) \right] &= 
		 \frac{(2 - u^2) D_x}{2(1-u^2) }  + \frac{u^2 D_y }{2(1-u^2) } \,,\\
		\mathbb{E}_{x,y}\left[ y^2(t) \right] 
		&= \frac{u^2 D_x }{2(1-u^2)}  + \frac{(2 - u^2) D_y }{2(1-u^2)} \,,\\
		\mathbb{E}_{x,y}\left[ x(t)y(t) \right] 
		&= \frac{u (D_x + D_y)}{2(1-u^2) }   \,.	
	\end{split}
\end{equation}
In the limit when the components are decoupled, i.e., when $u = 0$, (such that one has two independent generalized Ornstein-Uhlenbeck processes driven by dichotomous noises), one finds from eqs. \eqref{I} 
that 
\begin{equation}
\mathbb{E}_{x}\left[ x^2(t) \right] = D_x \frac{2\lambda_x}{1+ 2\lambda_x} \,, \quad 	\mathbb{E}_{y}\left[ y^2(t) \right] = D_y \frac{ 2\lambda_y}{1+2 \lambda_y} \,,	\quad  \mathbb{E}_{x,y}\left[ x(t) y(t)\right] = 0 \,,
		\end{equation}

 in the limit $t =  \infty$.

\begin{figure}[ht]
	\begin{center}
		\includegraphics[scale=0.5]{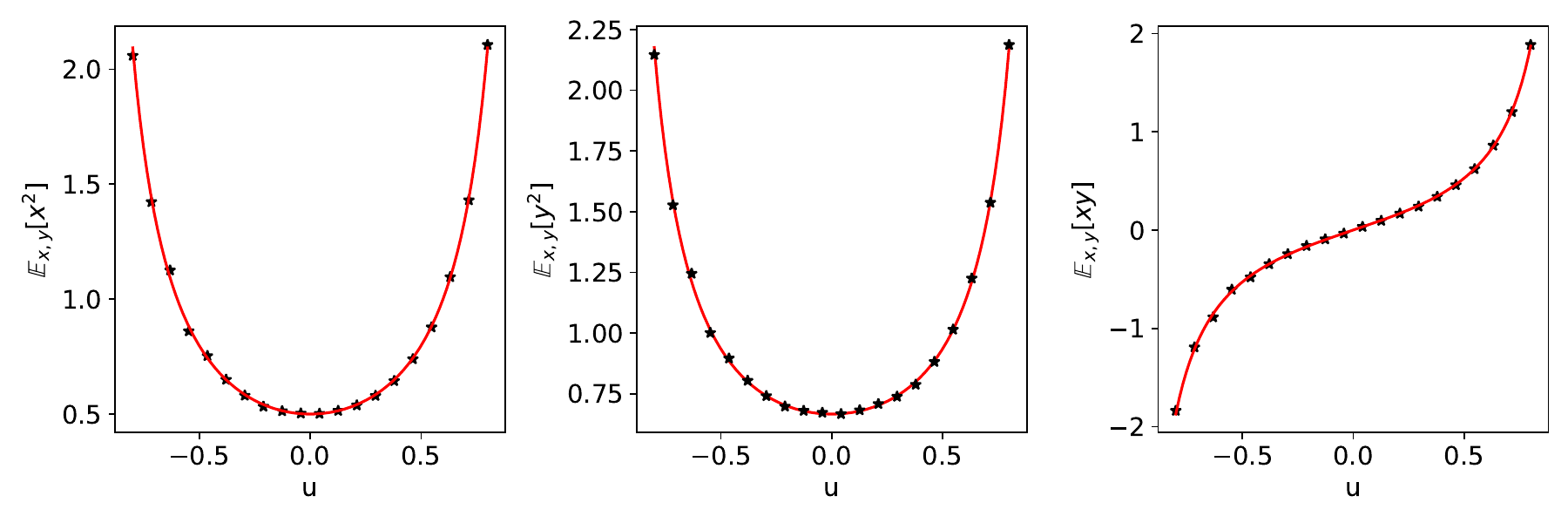}
	\end{center}
	\caption{The variances (second moments) of the $x$- and $y$-components in the limit $t \to \infty$ (left and middle panels), as well as the cross-moment $x(t) y(t)$ (right panel) as functions of the coupling parameter $u$ for $\lambda_x=0.5$, $\lambda_y=2.5$, $v_x=1.0$, and $v_y=2$. The solid curves correspond to the exact analytical expressions, while the stars indicate the numerical results.} \label{fig:moments}
\end{figure}

In Fig. \ref{fig:moments} we depict the variances (the second moments) of the $x$- and $y$-components in the limit $t \to \infty$ (left and middle panels), as well as the cross-moment 
  $xy$ (right panel) as functions of the coupling parameter $u$. We observe that both variances are non-monotonic functions of the coupling parameter and are minimal in the decoupled case, i.e., $u = 0$, which means that the coupling effectively increases the fluctuations of values of the SG components, and diverge when $u \to \pm 1$. The cross-moment is an antisymmetric function of $u$: for negative $u$ the components are anti-correlated, while for positive coupling parameter the components are positively correlated. The cross-moment also diverges when $u \to \pm 1$.

\subsection{Specific angular momentum}

The specific angular momentum of the SG is defined in eq. \eqref{L}.
Multiplying the first equation in eqs. \eqref{LE} by $y(t)$ and the second one - by $x(t)$, we can formally rewrite
eq. \eqref{L} as
\begin{equation}
	\label{L2}
	L = u \left(x^2(t) - y^2(t) \right) + x(t) \zeta_y(t) - y(t) \zeta_x(t) \,.
\end{equation} 
One notices that the instantaneous value of the specific angular moment
is the sum of two contributions: 
the difference of the squared components, and the difference of products $x(t) \zeta_y(t)$ and $y(t) \zeta_x(t)$.
Contrary to the standard BG model (see, e.g., \cite{Dotsenko2024}), the instantaneous values of dichotomous noises at time instant $t$ are not statistically-independent of the instantaneous positions of the components and hence, they contribute to the mean value of the angular momentum. We find that in the limit  $t \to \infty$ this contribution reads 
\begin{equation}\label{eq:noise}
		\mathbb{E}_{x,y}\left[ x(t) \zeta_y(t) -y(t) \zeta_x(t) \right] 
		=D_y\frac{2u \lambda_y}{(1 + 2\lambda_y)^2-u^2}-D_x\frac{2 u \lambda_x}{(1 + 2\lambda_x)^2-u^2} \,.
\end{equation}

Note that the above mean value is exactly equal to zero for the standard BG model. Interestingly enough, the contribution to $L$ in eq. \eqref{L2} due to correlations between the noises and the instantaneous positions has always a different sign as compared to the term $ u \left(x^2(t) - y^2(t) \right)$ in eq. \eqref{L2}, and therefore, this contribution  always \textit{diminishes} the absolute value of the mean angular momentum. 
Combining eqs.\ref{eq:noise} and \eqref{momentsxy}, we arrive at the following result for the first moment of the angular momentum of the SG in the limit $t \to \infty$:

\begin{align}\label{eq:L}
	\mathbb{E}_{x,y}\left[ L \right]& = u\left(D_x\frac{4\lambda_x^2}{(1+2\lambda_x)^2 - u^2}-D_y\frac{4\lambda_y^2}{(1+2\lambda_y)^2 - u^2 }\right) \,.
\end{align}
The above formula generalizes the result obtained for the standard BG model, which simply follows from eq. \eqref{eq:L} by setting $\lambda_x = \lambda_y = \infty$ to give
\begin{align}
	\label{LWGN}
	\mathbb{E}^{(GWN)}_{x,y}\left[ L \right]& = u\left(D_x-D_y\right) \,.
	\end{align}
Equation \eqref{eq:L} has a very clear structure which permits to  
establish the conditions under which the specific angular momentum, (and hence, the torque), has a non-zero value such that the SG subject to dichotomous noises has a tendency to perform a stochastic rotational motion. We conclude that the mean specific angular momentum is not equal to zero when necessarily 
the components have a non-zero coupling, $u \neq 0$, and either non-equal switching rates, $\lambda_x \neq \lambda_y$ or non-equal amplitudes of the dichotomous noises, $v_x \neq v_y$. Therefore, the behavior appears somewhat similar to that in the BG model driven by two statistically-independent \textit{fractional} Gaussian noises, for which the mean specific angular momentum is not equal to zero when $u \neq 0$ and either the respective amplitudes or the Hurst exponents characterizing the noises are not equal to each other \cite{Squarcini2022b}.

\begin{figure}[H]
		\includegraphics[scale=0.35]{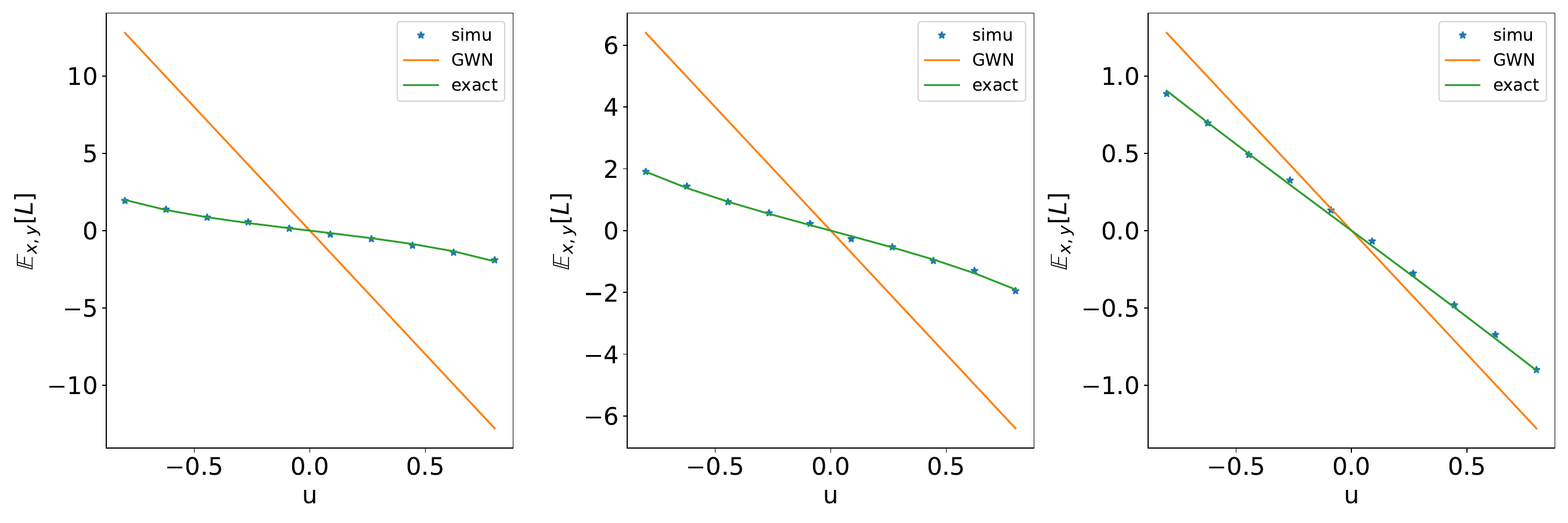}
	\caption{The mean specific angular momentum $\mathbb{E}_{x,y}\left[ L \right] $, (eq.\eqref{eq:L}, green curve) as function of $u$ for $\lambda_x =\lambda_{y}=0.25$ (left panel), $\lambda_x =\lambda_{y}=0.5$ (middle panel) and $\lambda_x =\lambda_{y}=2.5$ (right panel) with fixed non-equal amplitudes of noises, $v_x=1$ and $v_y=3$.  The magenta curve depicts the behavior of the mean specific angular momentum in the Gaussian white-noise (GWN) limit, (see eq. \eqref{LWGN} and  \cite{Viot2024}).
		Stars depict the results  of numerical simulations.} \label{fig:Ldicho}
\end{figure}

Figure \ref{fig:Ldicho} presents the mean specific angular momentum of the SG as function of the coupling parameter $u$ for equal switching rates $\lambda_x=\lambda_y$ and different values of the amplitudes of noises $v_x$ and $v_y$. The magenta curve depicts the corresponding result in the Gaussian white-noise limit in eq. \eqref{LWGN}. We observe 
that for dichotomous noises the value of the  mean specific angular momentum, and hence, of the torque exerted on the particle is always smaller than the white-noise result and tends to it, as it should,  upon a gradual increase of the switching rates.

\subsection{Specific angular velocity}
\label{angularvel}

Calculation of the mean specific angular velocity, its higher moments, as well of its PDF $P(W)$ necessitates the knowledge of the full position PDF, which goes beyond the scope of the present paper. Instead, here we resort to a numerical analysis of the random variable $W$ in eq. \eqref{W}. To this end, we solve numerically eqs. \eqref{LE} for given realizations of noises and generate sufficiently long trajectories $x(t)$ and $y(t)$, which permits us to get the instantaneous values of the specific angular momentum and of the specific angular velocity. Then, we perform averaging over $100$ realizations of trajectories.

\begin{figure}[h!]
	\begin{center}
\includegraphics[scale=0.35]{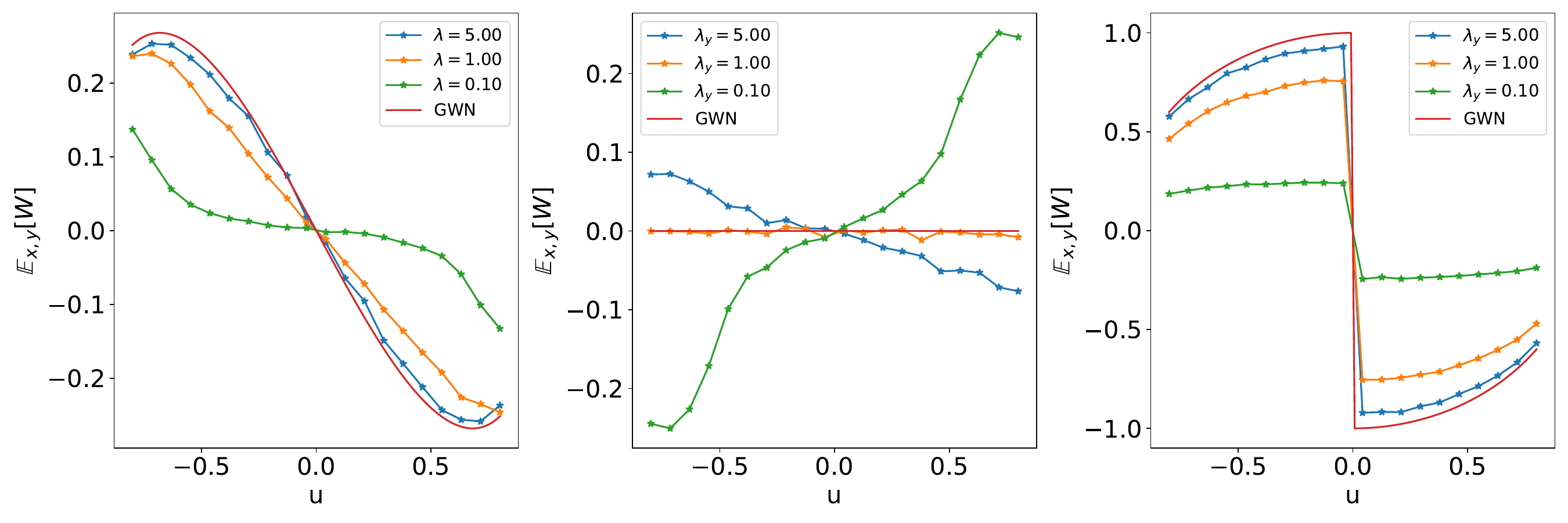}
	\end{center}
	\caption{Numerically-evaluated mean specific angular velocity $\mathbb{E}_{x,y} [W]$ as function of  $u$. Left panel: $\lambda = \lambda_x = \lambda_y =5$, $1$ and $0.1$ (see the inset) with fixed amplitudes of noises $v_x=1$ and $v_y=3$. The red solid 
		curve depicts the exact expression for the Gaussian white-noise (see the expression in, e.g., \cite{Viot2024}). Other curves are merely guides to the eye obtained by a freehand interpolation of symbols depicting the numerical results. Middle panel: equal amplitudes of noises $v_x = v_y = 3$, $\lambda_x = 1$ and variable $\lambda_y$ (see the inset). Right panel: fixed ratios $D_x =0$,  $ D_y =1$, $\lambda_x = 1$  and variable $\lambda_y$. The solid red curves depicts the exact expression for the Gaussian white-noise.}\label{fig:angaularvel}
\end{figure}

In Fig. \ref{fig:angaularvel} we depict the behavior of the mean specific angular 
velocity of the SG as function of $u$ in three situations: In the left panel we consider unequal amplitudes of noises, $v_x = 1$ and $v_y =3$, and take equal switching rates - $\lambda_x  = \lambda_y = 0.1$, $1$ and $5$. We observe that except for the trivial decoupled case $u=0$, the mean angular velocity is not equal to zero. Likewise the mean angular momentum, $\mathbb{E}_{x,y} [W]$ is positive for $u < 0$ and negative - for $u > 0$. Further on, we see that for the largest value of the switching rates, i.e., $\lambda_x = \lambda_y = 5$, the mean angular velocity of the SG is close, for the whole range of variation of the coupling parameter,  to the  result  obtained for the Gaussian  white-noise (see the red solid line and \cite{Viot2024}). For smaller values of the switching rates we observe a significant departure from the behavior in the standard BG model and even an appearance of the inflection points.  In the middle panel we depict the behavior of $\mathbb{E}_{x,y} [W]$ for equal amplitudes of noises ($v_x = v_y = 3$), fixed switching rate $\lambda_x = 1.0$ and variable $\lambda_y  = 0.1$, $1.0$ and $5$. We observe that the mean specific angular velocity vanishes for $\lambda_y = 0.5$, as it should because here both amplitudes of noises and the switching rates are equal, and is not equal to zero for two other cases. Interestingly enough, the mean angular velocity of the SG changes its sign as function of $\lambda_y$: for $u < 0$ the mean angular velocity is positive for $\lambda_y = 5$ and negative for $\lambda_y = 0.1$. Lastly, in the right panel we depict  $\mathbb{E}_{x,y} [W]$ as function of the coupling parameter in case when we have  diffusion coefficients $D_x =0$, $D_y = 1$ (see the definition in eq. \eqref{diffD}), fixed switching rate of the $x$-component, $\lambda_x = 1$, and variable $\lambda_y$. A salient feature here is that the mean specific angular velocity exhibits a discontinuous behavior at $u = 0$. The discontinuity is most pronounced for the standard BG model but also persists for the  SG driven by dichotomous noises with finite switching rates, becoming less pronounced for smaller values of $\lambda_y$. 

\section{Probability density functions of $\mathcal{S} = x + y$ and $\mathcal{D} = x - y$ in the steady-state: General results}
\label{PDF}

In order to calculate $P(\mathcal{S})$ and $P(\mathcal{D})$ we seek to determine first the characteristic functions
\begin{align}
\begin{split}
\label{Ux}
&\Phi_{\mathcal{S}}(\omega) = \mathbb{E}_{x}\left[\exp\left(i \omega \mathcal{S}_x\right) \right]  \mathbb{E}_{y}\left[\exp\left(i \omega 
\mathcal{S}_y\right) \right] \\ &=  \mathbb{E}_{x}\left[\exp\left(i \omega \int^t_0 d\tau e^{-(1-u)(t - \tau)} \zeta_x(\tau)\right)\right] \,
 \mathbb{E}_{y}\left[\exp\left(i \omega \int^t_0 d\tau e^{-(1-u)(t - \tau)} \zeta_y(\tau)\right)\right] \,,
\end{split}
\end{align}
and
\begin{align}
\begin{split}
\label{Vx}
&\Phi_{\mathcal{D}}(\omega) = \mathbb{E}_{x}\left[\exp\left(i \omega \mathcal{D}_x\right) \right]  \mathbb{E}_{y}\left[\exp\left(- i \omega 
\mathcal{D}_y\right) \right] \\ &=  \mathbb{E}_{x}\left[\exp\left(i \omega \int^t_0 d\tau \, e^{-(1+u)(t - \tau)} \zeta_x(\tau)\right)\right] \,
 \mathbb{E}_{y}\left[\exp\left(- i \omega \int^t_0 d\tau \, e^{-(1+u)(t - \tau)} \zeta_y(\tau)\right)\right] \,,
\end{split}
\end{align}
with $\mathcal{S}_a$ and $\mathcal{D}_a$, $a = x,y$, defined in eqs. \eqref{inst}. 

Here, a trivial observation is in order: since the noises are symmetric, each multiplier in eqs. \eqref{Ux} and \eqref{Vx} is an even function of $\omega$. This signifies that the $"-"$ sign in the last multiplier in eq. \eqref{Vx} is irrelevant and 
 the characteristic functions, hence, the PDFs of $\mathcal{S}$ and $\mathcal{D}$ become identical upon a mere replacement of $u$ by $-u$. Hence, it suffices to calculate either of them; say, $\Phi_{\mathcal{S}}(\omega)$ and $P(\mathcal{S})$.

The exact result for the PDF $P(\mathcal{S}_x)$ can be deduced from the remarkable paper \cite{Sancho1984} (see also \cite{Balakrishnan1993}), in which the position PDF of the generalized Ornstein-Uhlenbeck process driven by a dichotomous noise has been studied. In our notations, the main result of \cite{Sancho1984} in the limit $t \to \infty$ is given by
\begin{align}
\label{sancho}
P(\mathcal{S}_x) = \frac{(1-u) \Gamma(1/2+\alpha_x)}{\sqrt{\pi} \Gamma(\alpha_x) v_x^{2 \alpha_x -  1}}
\begin{cases}
\left(v_x^2 - (1 - u)^2 \mathcal{S}_x^2\right)^{\alpha_x -  1},  \,\,\,  \quad |\mathcal{S}_x| \leq v_x/(1 - u)\\
0, \qquad  \qquad \qquad \,\,\, \,\,\,\,\,\,\,\,\,\,\,\,\,\,\, \, \quad |\mathcal{S}_x| > v_x/(1 - u) \,, 
\end{cases}
\end{align}
where  
\begin{align}
\label{alpha}
\alpha_x = \frac{\lambda_x}{(1-u)} .
\end{align}
For completeness, we present a streamlined derivation of the result in eqs. \eqref{sancho} and \eqref{alpha} in \ref{A}.

We dwell some more on the expression \eqref{sancho}. First, this PDF has a finite support $(-v_x/(1 - u),v_x/(1 - u))$. This is, of course, expected in light of the bounds imposed on the components 
$x(t)$ 
and 
$y(t)$
of the trajectories, as given in eq. \eqref{bounded}. By following the same reasoning that led to eq. \eqref{bounded}, we find that 
\begin{align}
	- \frac{v_x}{1-u} \left(1 - e^{-(1-u) t}\right) \leq \mathcal{S} \leq \frac{v_x}{1-u} \left(1 - e^{-(1-u) t}\right) \,,
	\end{align}
which yields, in the limit $t \to \infty$, the edges of the support in eq. \eqref{sancho}.
Second, one notices that $\alpha_x$ in eq. \eqref{alpha} is the key parameter controlling the shape  of $P(\mathcal{S}_x)$: for $\alpha_x < 1$ the probability density function is bimodal - it has a minimum at $\mathcal{S}_x=0$ and diverges at the boundaries of the support. Conversely, for $\alpha_x > 1$ the probability density function 
is unimodal - it has a maximum at $\mathcal{S}_x=0$ and vanishes at the edges of the support. In the borderline case $\alpha_x = 1$ the PDF is uniform.

Further on, we obtain
\begin{align}
\begin{split}
\label{k3}
&\mathbb{E}_{x}\left[\exp\left(i \omega \mathcal{S}_x\right) \right]  = \int^{v_x/(1-u)}_{-v_x/(1-u)} d\mathcal{S}_x \, e^{i \omega \mathcal{S}_x} \, P(\mathcal{S}_x)\\
& = 2^{\alpha_x - 1/2} \Gamma(1/2 + \alpha_x) \left(\frac{v_x}{1-u} \omega\right)^{1/2-\alpha_x} J_{\alpha_x - 1/2}\left(\frac{v_x}{1-u} \omega\right) \,,
\end{split}
\end{align}
where $J_{\nu}(\ldots)$ is the Bessel function of the first kind. 
In a similar way, we determine the second multiplier in eq. \eqref{Ux}, cf. eq. \eqref{k3}, 
\begin{align}
&\mathbb{E}_{y}\left[\exp\left(i \omega \mathcal{S}_y\right) \right]  = 2^{\alpha_y - 1/2} \Gamma(1/2 + \alpha_y) \left(\frac{v_y}{1-u} \omega\right)^{1/2-\alpha_y} J_{\alpha_y - 1/2}\left(\frac{v_y}{1-u} \omega\right) \,,
\end{align}
where 
\begin{align}
\alpha_y = \frac{\lambda_y}{1-u} \,.
\end{align}
Consequently, the characteristic function $\Phi_{\mathcal{S}}(\omega) $ is explicitly given by
\begin{align}
\begin{split}
\label{charx}
\Phi_{\mathcal{S}}(\omega)  &= \frac{2^{\alpha_x + \alpha_y - 1} \Gamma(1/2 + \alpha_x) \Gamma(1/2 + \alpha_y) v_x^{1/2-\alpha_x} v_y^{1/2-\alpha_y}}{(1 - u)^{1 - \alpha_x - \alpha_y}}  \\
& \times  \dfrac{J_{\alpha_x - 1/2}\left(\dfrac{v_x}{1 - u} \omega\right)}{\omega^{\alpha_x - 1/2}} \, \dfrac{J_{\alpha_y - 1/2}\left(\dfrac{v_y}{1 - u} \omega\right)}{\omega^{\alpha_y - 1/2}} \,.
\end{split}
\end{align}
Inverting the Fourier transform in the expression \eqref{charx}, we find that the PDF of 
$\mathcal{S}$ obeys
\begin{align}
\begin{split}
\label{pU}
P(\mathcal{S}) &= \frac{2^{\alpha_x + \alpha_y} \Gamma(1/2 + \alpha_x) \Gamma(1/2 + \alpha_y)}{2 \pi (1 - u)^{1 - \alpha_x - \alpha_y} } v_x^{1/2-\alpha_x} v_y^{1/2 - \alpha_y}  \\
& \times \int^{\infty}_0 d\omega \, \dfrac{J_{\alpha_x - 1/2}\left(\dfrac{v_x}{1 - u} \omega\right)}{\omega^{\alpha_x - 1/2}} \, \dfrac{J_{\alpha_y - 1/2}\left(\dfrac{v_y}{1 - u} \omega\right)}{\omega^{\alpha_y - 1/2}} \, \cos(\omega \mathcal{S})  \,,
\end{split}
\end{align}
which represents our main general result. 

The integral in eq. \eqref{pU} cannot be performed exactly for arbitrary values of parameters. However, several general  
statements can be made: \\
-- The probability density function $P(\mathcal{S})$ (and hence, $P(\mathcal{D})$) has a finite support (cf. eq. \eqref{sancho})
\begin{align}
\label{support}
\mathcal{S} \in (-\Sigma,\Sigma) \,, \qquad \Sigma = \frac{v_x + v_y}{1-u}  \,.
\end{align}
We will discuss the behavior at the edges of the support below, in Subsec. \ref{edges} and Sec. \ref{part}.\\
-- This probability density function exhibits an \textit{irregular} behavior at
\begin{align}
\label{Omega}
\mathcal{S} = \pm \Omega \,, \qquad \Omega = \frac{|v_x - v_y|}{1 - u} \,.
\end{align}
This irregularity, depending on the case in hand, can be either a discontinuity of the  function itself or 
of its derivative (see Subsec. \ref{edges} and Sec. \ref{part}). Note that for the random variable $\mathcal{D} = x -  y$ of eq. \eqref{SD}
one has an irregular behavior at $\mathcal{D} = \pm \Omega'$ with $\Omega' = |v_x - v_y|/(1+u)$, and $\mathcal{D} \in (-\Sigma',\Sigma')$ with $\Sigma' = (v_x + v_y)/(1+u)$. \\
-- Recalling that $P(\mathcal{S}_x)$ in eq. \eqref{sancho} exhibits three different kinds of behavior depending whether $\alpha_x$ is less than $1$, is equal to $1$ or is greater than $1$, we may conclude that also $P(\mathcal{S})$ in eq. \eqref{pU} will show a different behavior 
whether
\begin{flushleft}
A) $\alpha_x < 1$ and $\alpha_y < 1$\\
B) $\alpha_x < 1$ and $\alpha_y = 1$, or vice versa\\
C) $\alpha_x  < 1$ and $\alpha_y > 1$, or vice versa\\
D) $\alpha_x = \alpha_y = 1$\\
E) $\alpha_x = 1$ and $\alpha_y > 1$, or vice versa\\
F) $\alpha_x > 1$ and $\alpha_y > 1$\\
\end{flushleft}
Moreover, in case of unequal $\alpha_x$ and $\alpha_y$ one may also expect to observe different shapes of $P(\mathcal{S})$ depending whether $v_x$ is greater than $v_y$, or if $v_x < v_y$. 
Below, in Sec. \ref{part} we will illustrate the behavior of $P(\mathcal{S})$ for all 
these cases taking particular values of $\alpha_x$ and $\alpha_y$ for which the integral in eq. \eqref{pU}
can be evaluated; Namely, this is the case when  $\alpha_x$ and $\alpha_y$ are either half-integers (in which case $P(\mathcal{S})$ can be expressed through a combination of elliptic integrals) or when both are
 integers (in which case $P(\mathcal{S})$ is given as a combination of elementary functions and the inverse trigonometric functions). Note that the integral can be also performed exactly when $\alpha_x = \alpha_y$ and $v_x = v_y$, which is not an interesting case for us, because the specific angular momentum and the specific angular velocity vanish in this limit.
 
Before we proceed to Sec. \ref{part} presenting 
such a case-by-case analysis of the possible functional forms of $P(\mathcal{S})$, we discuss below several other general results, including the derivation of the moments of $P(\mathcal{S})$ (and of $P(\mathcal{D})$) or arbitrary order, behavior of $P(\mathcal{S})$ in the white-noise limit, its behavior at $\mathcal{S}=0$,  as well as at the irregular points $\mathcal{S} = \pm \Omega$ and at the edges of the support, $\mathcal{S} = \pm \Sigma$.
 
\subsection{Moments of $P(\mathcal{S})$}

Note that the moments of $\mathcal{S}$ of arbitrary order can be obtained from eq. \eqref{charx} very directly. To this end, we take advantage of the well-known Taylor series expansion (see \cite{gradshteyn2007}, Section 8.442)
\begin{align}
	\begin{split}
		\label{ser}
		\dfrac{J_{\nu}\left(a \omega\right)}{\omega^{\nu}} \, \dfrac{J_{\mu}\left(b \omega\right)}{\omega^{\mu}} &= \frac{a^{\nu} b^{\mu}}{2^{\nu+\mu} \Gamma(\mu + 1)} \sum_{k=0}^{\infty} \frac{(-1)^k a^{2 k}}{4^k \, k! \, \Gamma(\nu + k + 1)} \\&\times \,_2F_1\left(-k, - \nu - k; \mu + 1; \frac{b^2}{a^2}\right) \, \omega^{2 k} \,, \quad b \leq a \,, 
	\end{split}
\end{align} 
where $\,_2F_1(\ldots)$ is the Gauss hypergeometric function, and an analogous expression with $\nu$ replaced by $\mu$ (and vice versa) for $b \geq a$. This permits us to find the Taylor expansion of the characteristic function in eq. \eqref{charx}  in  powers of $\omega$ and hence, to determine the moments of $\mathcal{S}$. 
Recalling next the definition of the Jacobi polynomials $P_n^{(p,q)}(\xi)$ (with $\xi \in (-1,1)$) in terms of the hypergeometric function (see \cite{gradshteyn2007}, Section 8.962), we find that for $v_x \geq v_y$ and arbitrary $\lambda_x$ and $\lambda_y$, the moments of $\mathcal{S}$ of an even order are 
\begin{align}
	\begin{split}
		\label{mom1}
		\mathbb{E}_{x,y}\left[\mathcal{S}^{2 n}\right] &=  \mathbb{E}_{x,y}\left[\left(x + y\right)^{2 n}\right] = (2 n)! \frac{ \Gamma(1/2 + \alpha_x) \Gamma(1/2 + \alpha_y) }{ \Gamma(1/2 + \alpha_x + n) \Gamma(1/2 + \alpha_y + n)} \\& \times \left(\frac{v_x}{2 (1 - u)}\right)^{2 n} \, P_n^{(p,q)}\left(1 - \frac{2 v_y^2}{v_x^2}\right) \,, \quad p = \alpha_y - 1/2 \,, q = - 2n - \alpha_x - \alpha_y \,,
	\end{split}
\end{align}
while the moments of an odd order are all identically equal to zero. Respectively, for $v_x \leq v_y$ we have
\begin{align}
	\begin{split}
		\label{mom2}
		\mathbb{E}_{x,y}\left[\mathcal{S}^{2 n}\right] &=   (2 n)! \frac{ \Gamma(1/2 + \alpha_x) \Gamma(1/2 + \alpha_y) }{ \Gamma(1/2 + \alpha_x + n) \Gamma(1/2 + \alpha_y + n)} \\& \times \left(\frac{v_y}{2 (1 - u)}\right)^{2 n} \, P_n^{(p,q)}\left(1 - \frac{2 v_x^2}{v_y^2}\right) \,, \quad p = \alpha_x - 1/2 \,, q = - 2n - \alpha_x - \alpha_y \,.
	\end{split}
\end{align}
The moments of $\mathcal{D}$ follow from eqs. \eqref{mom1} and \eqref{mom2} by a mere change of the sign of the coupling parameter $u$. 
 
\subsection{White-noise limit: Standard BG model}

To set up the scene and to highlight a wealth of the behavior in case of the SG driven by the dichotomous noises, which will be discussed later in this Section,  we start with the analysis of the PDF in eq. \eqref{sancho} in the limit $\lambda_{a} \to \infty$ ($\tau_{a} \to 0$), $a = x,y$, and $v_{a} \to \infty$ with the ratios $D_{a}  = v_{a}^2/2 \lambda_{a}$ in eq. \eqref{diffD} being kept fixed. In this limit, the dichotomous noises 
in eqs. \eqref{noise1} and \eqref{noise2} become Gaussian white-noises and the expression \eqref{sancho} for the PDF of the $x$-component attains the form
\begin{align}
P(\mathcal{S}_x) = \sqrt{\frac{1-u}{2 \pi D_x}} \exp\left(- \frac{1 - u}{2 D_x} \mathcal{S}_x^2\right) \,, 
\end{align}
where $\mathcal{S}_x \in (-\infty,\infty)$ because $v_x$ is infinitely large in this limit.  In a similar way, we find the PDF of $\mathcal{S}_y$ and eventually, the probability density function of $\mathcal{S}$: 
\begin{align}
\label{Gauss}
P(\mathcal{S}) = \sqrt{\frac{(1-u)}{2 \pi (D_x + D_y)}} \exp\left(- \frac{1 - u}{2 (D_x + D_y)} \mathcal{S}^2\right) \,, \quad S \in (-\infty,\infty) \,,
\end{align}
whose moments of an odd order are equal to zero, while the moments of an even order obey, quite trivially,
\begin{align}
\label{mom3}
\mathbb{E}_{x,y}\left[\mathcal{S}^{2 n}\right] &= \frac{\Gamma(n+1/2)}{\sqrt{\pi}} \left(\frac{2 (D_x + D_y)}{(1 - u)}\right)^n \,,
\end{align}
and have, of course, much simpler form than the moments in eqs. \eqref{mom1} and \eqref{mom2} for the dichotomous noises with finite $\lambda_{a}$ and $v_{a}$. 
We also note parenthetically that for the standard BG model all the position PDFs - not only the marginal ones but also the joint position PDF $P(x,y)$ are Gaussian functions for all times, albeit the coefficients in front of $x^2$, $y^2$ and $x y$ are quite complicated functions of the system's parameters and time. 

\begin{figure}[h!]
\centering
\includegraphics[width=60mm]{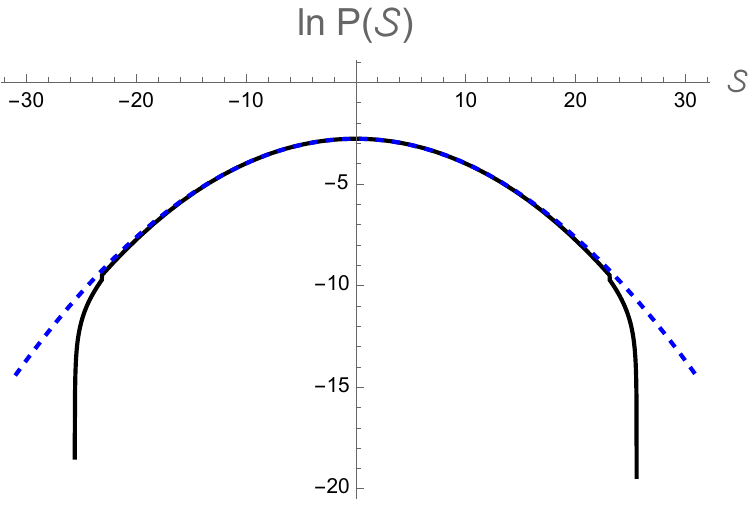}  
\caption{Logarithms of the probability density functions $P(\mathcal{S})$ in eq. \eqref{pU}  (black solid curve) and the white-noise-limit asymptotic form in eq. \eqref{Gauss} (blue dashed curve)
as functions of $\mathcal{S}$
for $u=3/4$, $v_x = v_y = 10$ and $\lambda_x = \lambda_y = 9.7$.   
 }
\label{fig:F2} 
\end{figure} 

We depict the PDF of eq. \eqref{Gauss} in Fig. \ref{fig:F2} confronting it against the exact form of the probability density function in eq.  \eqref{pU}, which is computed numerically in Mathematica. We observe that even for quite moderate values of $\lambda_a$ and $v_a$, (with, however, $\lambda_x$ and $\lambda_y$ exceeding some critical values - see the case F above), the Gaussian form in eq. \eqref{Gauss} approximates quite well the central part  of $P(\mathcal{S})$ in eq. \eqref{pU} away from the edges of the support. Behavior in case of sufficiently small
values of $\lambda_x$ and $\lambda_y$ will be, of course, markedly different.

\subsection{Behavior of $P(\mathcal{S})$ of eq. \eqref{pU} at $\mathcal{S}=0$.}

Consider next the behavior at $\mathcal{S} = x + y = 0$, in which case the expression \eqref{pU} is 
the discontinuous Weber-Shaftheitlin integral (see \cite{Bateman2023}, volume II, Section 7.7.4).  By using the integral, we find that for arbitrary values of parameters
\begin{align}
\label{pU0}
P(\mathcal{S}= 0) = \frac{1 - u}{\sqrt{\pi}}
\begin{cases}
\frac{\Gamma(\alpha_y + 1/2)}{\Gamma(\alpha_y) \, v_y} \,_2F_1\left(1/2, 1 - \alpha_y; \alpha_x +1/2; v_x^2/v_y^2\right),   \quad  \quad  v_x \leq v_y,\\
\frac{\Gamma(\alpha_x + 1/2)}{\Gamma(\alpha_x) \, v_x} \,_2F_1\left(1/2, 1 - \alpha_x; \alpha_y +1/2; v_y^2/v_x^2\right),   \quad  \quad v_x \geq v_y . \\
\end{cases}
\end{align}
\begin{figure}[h!]
\centering
\includegraphics[width=60mm]{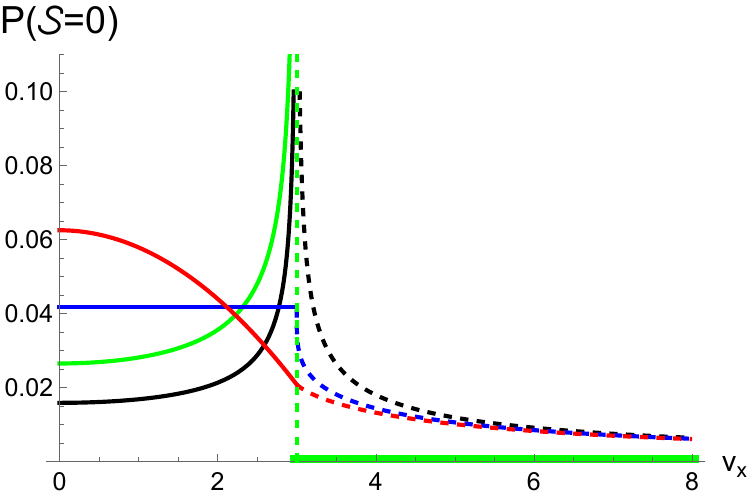}  
\includegraphics[width=60mm]{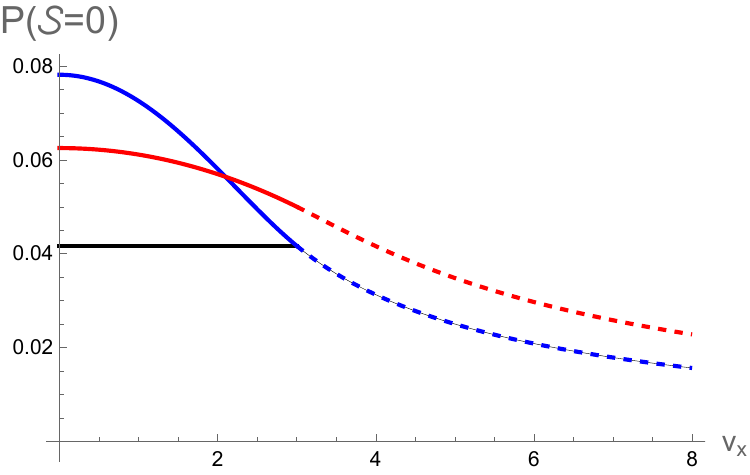}
\caption{$P(\mathcal{S} = x + y = 0)$ in eq. \eqref{pU0}  
as function of $v_x$
for $u=3/4$ and $v_y = 3$. Solid curves correspond to the expression in the first line in eq. \eqref{pU0}, while the dashed ones - to the second line in this expression. Left panel: Green curve depicts the behavior of $P(\mathcal{S}=0)$ in the special situation with $\alpha_x = 0$ and $\alpha_y = 1/4$ (this is a particular instance of the case A, which will be considered in more detail in Sec. \ref{Zx0}), black curve - for $\alpha_x = 1/4$ and $\alpha_y = 1/4$ (case A), blue curve  - for $\alpha_x = 1/4$ and $\alpha_y = 1$ (case B) and 
red curve - for $\alpha_x = 1/4$ and $\alpha_y = 2$ (case C). Right panel: Black curve depicts $P(\mathcal{S}=0)$ for $\alpha_x = \alpha_y=1$ (case D), blue curve - for $\alpha_x=1$ and $\alpha_y = 3$ (case E), while the red curve corresponds to the case F with $\alpha_x = \alpha_y = 2$.
}
\label{fig:01} 
\end{figure} 

In Fig. \ref{fig:01} we depict $P(\mathcal{S}=0)$ defined in eqs. \eqref{pU0} as function of $v_x$ for fixed $v_y = 3$, $u=3/4$ and several choices of values of  $\alpha_x$ and $\alpha_y$, covering all possible cases, from A to F. On the left panel we plot $P(\mathcal{S}=0)$ for the cases A, B and C. We observe that in the case $A$ the PDF $P(\mathcal{S}=0)$ is a discontinuous function of $v_x$ which diverges as a power-law when $v_x$ approaches $v_y$ from below. This is a generic behavior in this case which we illustrate by choosing $\alpha_y = 1/4$ and
the two choices of $\alpha_x$: $\alpha_x = 0$, (i.e., the function $\zeta_x(t)$ is not a noise here and is not fluctuating with time maintaining its value at $t=0$, see the solid green curve) and  
$\alpha_x = 1/4$ (both $\zeta_x(t)$ and $\zeta_y(t)$ are dichotomous noises, see the solid black curve). However, for these two situations the behavior past $v_x = v_y$ appears to be quite different. For the latter case $P(\mathcal{S}=0)$ is greater than zero for any value of $v_x > v_y$, and diverges as a power-law when $v_x \to v_y$ from above (see the dashed black curve). For the former case we have $P(\mathcal{S}=0) \equiv 0$ for any $v_x  > v_y$ (see the thick green horizontal line). We note parenthetically that in this situation we may expect a very different behavior of the PDF $P(W)$ of the angular velocity $W$, as compared to the behavior  in the standard BG model; namely, the power-law tails $P(W) \sim 1/|W|^3$ should be absent. As mentioned in the Introduction, for the BG model 
such tails emerge due to the fact that the BG most probably resides at the origin of the coordinate axes such that the moment of inertia is typically equal to zero and $W$ has therefore an infinitely large value (see eq. \eqref{W}).  For the SG driven by dichotomous noises with  $\alpha_x = 0$, $0 <  \alpha_y < 1$ and $v_x > v_y$ the probability of finding the SG at the origin is exactly equal to zero, as evidenced in Fig. \ref{fig:01}, and the moment of inertia should therefore be strictly greater than zero such that $W$ should, in principle,  possess the moments of arbitrary order. In the Sec. \ref{Zx0} we will return to this particular situation and discuss in detail the form of the function $P(\mathcal{S})$, (as well as its behavior for $v_x \to 0$), and will show that, in fact, there is a finite interval of values of $\mathcal{S}$ for which $P(\mathcal{S}) \equiv 0$. 
The gyration characteristics in this particular case will be studied elsewhere \cite{Herbeau2026}. 
In the case B, as exemplified here by the choice $\alpha_x=1/4$ and $\alpha_y =1$, the PDF $P(\mathcal{S}=0)$ appears to be a piecewise continuous function of $v_x$ (see solid and dashed blue curves): it is $v_x$-independent constant for $v_x < v_y$, exhibits a cusp at $v_x = v_y$ (such that its derivative has a finite jump at this point) and is a monotonically decreasing function of $v_x$ for $v_x > v_y$. Lastly, in the case C for which we take as the representative example the choice
$\alpha_x=1/4$ and $\alpha_y = 2$, $P(\mathcal{S}=0)$ is a monotonically decreasing peace-wise continuous function of $v_x$ whose derivative shows a discontinuity for $v_x = v_y$.
In the right panel in Fig. \ref{fig:01} we depict the behavior corresponding to the regimes D to F. We observe that $P(\mathcal{S}=0)$ is piecewise continuous with a cusp at $v_x = v_y$ only in the case D when $\alpha_x = \alpha_y = 1$ (see the black solid and dashed curves). For the cases E (here, $\alpha_x = 1$ and $\alpha_y = 3$, blue solid and dashed curves) and F (here, $\alpha_x = \alpha_y = 2$, red solid and dashed curves) this function and its derivative are continuous functions of $v_x$. Note that in the large-$v_x$ limit and $\alpha_x > 0$, the function $P(\mathcal{S}=0)$ shows the universal asymptotic behavior $P(\mathcal{S}=0) \sim (1 - u)/v_x$, 
which trivially follows from the expression in the second line in eq. \eqref{pU0}.

\subsection{The probability density function $P(\mathcal{S})$ at $\mathcal{S} = \pm \Omega$ and $\mathcal{S} = \pm \Sigma$}
\label{edges}

Here we are concerned with the behavior of $P(\mathcal{S})$ at the irregular points $\mathcal{S} = \pm \Omega$, (see eq. \eqref{Omega}), and also at the edges of the support, $\mathcal{S} = \pm \Sigma$, (see eq. \eqref{support}). Before we proceed, it is worthy to note that this behavior can be quite non-trivial. Indeed, consider first the limit $\alpha_x \to 0$ and $\alpha_y \to 0$ (we will consider the special case $\alpha_x \to 0$ and $\alpha_y > 0$ in more detail in Sec. \ref{Zx0}), in which $P(\mathcal{S})$ becomes a sum of four delta-functions. In other word, $P(\mathcal{S} = \pm \Omega)$ and $P(\mathcal{S} = \pm \Sigma)$ diverge in this special case. 
Second, note that $\Omega \equiv 0$ 
when $v_x = v_y$, such that here $P(\mathcal{S} = \pm \Omega = 0)$ is simply given by expression \eqref{pU0}, for arbitrary values of $\alpha_x$ and $\alpha_y$. In the previous subsection we have discussed the behavior of this function and have shown that it can be quite different; namely, depending on the values of $\alpha_x$ and $\alpha_y$  it may diverge or stay finite. Below we inquire for which values of $\alpha_x$ and $\alpha_y$, the values of the PDF at the irregular points,  $P(\mathcal{S} = \pm \Omega)$, and at the edges of the support, $P(\mathcal{S} = \pm \Sigma)$, remain finite or are infinitely large.

In quest for the answer, we first represent
\begin{align}
\begin{split}
\label{qs}
P(\mathcal{S}= \pm \Omega) & = \int^{\infty}_0 d\omega \Big(F_c\left(v_x,\alpha_x,\omega\right) F_c\left(v_y,\alpha_y,\omega\right) + F_s\left(v_x,\alpha_x,\omega\right) F_s\left(v_y,\alpha_y,\omega\right) \Big) \,, \\
P(\mathcal{S}= \pm \Sigma) & =  \int^{\infty}_0 d\omega \Big(F_c\left(v_x,\alpha_x,\omega\right) F_c\left(v_y,\alpha_y,\omega\right) - F_s\left(v_x,\alpha_x,\omega\right) F_s\left(v_y,\alpha_y,\omega\right) \Big) \,,
\end{split}
\end{align}
 where 
 \begin{align}
 \begin{split}
 F_c(v,\alpha,\omega) & = \frac{2^{\alpha} \Gamma(1/2+ \alpha) v^{1/2-\alpha}}{\sqrt{2 \pi} (1 - u)^{1/2 - \alpha}} \dfrac{J_{\alpha-1/2}\left(\dfrac{v}{1-u} \omega\right)}{\omega^{\alpha-1/2}}  \cos\left(\frac{v}{1 - u} \omega \right) \,,\\
 F_s(v,\alpha,\omega) &= \frac{2^{\alpha} \Gamma(1/2+ \alpha) v^{1/2-\alpha}}{\sqrt{2 \pi} (1 - u)^{1/2 - \alpha}} \dfrac{J_{\alpha-1/2}\left(\dfrac{v}{1-u} \omega\right)}{\omega^{\alpha-1/2}}  \sin\left(\frac{v}{1 - u} \omega \right) \,.
 \end{split}
\end{align}
Next, we perform Laplace transforms of the eqs.  \eqref{qs} over the amplitudes $v_x$ and $v_y$ to get
\begin{align}
\begin{split}
\label{om}
\tilde{P}_{\Omega} &= \int^{\infty}_0 dv_x   \int^{\infty}_0 dv_y \, e^{-z_x v_x - z_y v_y} P(\mathcal{S}= \pm \Omega) \\
& =  \int^{\infty}_0 d\omega \Big(\tilde{F}_c\left(\alpha_x,\omega\right) \tilde{F}_c\left(\alpha_y,\omega\right) + \tilde{F}_s\left(\alpha_x\omega\right) \tilde{F}_s\left(\alpha_y,\omega\right)\Big) \,,\\
\tilde{P}_{\Sigma} &= \int^{\infty}_0 dv_x   \int^{\infty}_0 dv_y \, e^{-z_x v_x - z_y v_y} P(\mathcal{S}= \pm \Sigma) \\
& = \int^{\infty}_0 d\omega \Big(\tilde{F}_c\left(\alpha_x,\omega\right) \tilde{F}_c\left(\alpha_y,\omega\right) - \tilde{F}_s\left(\alpha_x\omega\right) \tilde{F}_s\left(\alpha_y,\omega\right)\Big) \,,\\
\end{split}
\end{align}
where
\begin{align}
\begin{split}
\tilde{F}_c\left(\alpha,\omega\right) &= \frac{1}{\sqrt{\pi} z} \,_3F_2\left(1, \frac{\alpha}{2}, \frac{\alpha+1}{2}; \alpha,  \alpha + \frac{1}{2}; - \frac{4 \omega^2}{(1-u)^2 z^2}\right) \,, \\
\tilde{F}_s\left(\alpha,\omega\right) &=  \frac{\omega}{\sqrt{\pi} (1-u) z^2} \,_3F_2\left(1, \frac{\alpha+1}{2}, \frac{\alpha}{2}+1;  \alpha + \frac{1}{2}, \alpha+1; - \frac{4 \omega^2}{(1-u)^2 z^2}\right)  \,,
\end{split}
\end{align}
where $_3F_2(\ldots)$ is the generalized hypergeometric function \cite{gradshteyn2007}. The kernels in the integrals over $\omega$ in eqs. \eqref{om}, i.e., 
\begin{align}
\label{kern}
\tilde{F}_c\left(\alpha_x,\omega\right) \tilde{F}_c\left(\alpha_y,\omega\right) \pm \tilde{F}_s\left(\alpha_x\omega\right) \tilde{F}_s\left(\alpha_y,\omega\right)
\end{align}
are analytic functions of $\omega$ in the limit $\omega \to 0$, i.e., are integrable. Consider next their behavior in the limit $\omega \to \infty$. We find that in this limit
\begin{align}
\begin{split}
\label{as1}
\tilde{F}_c\left(\alpha_x,\omega\right) = 
\begin{cases}
\left(c_1 \cos(\pi \alpha/2)/\omega^{\alpha}\right) \left(1 + O\left(1/\omega^{\alpha + 1}\right)\right),    \quad  \quad
 \alpha < 2 \,,  \\
c_2/\omega^2 + O\left(1/\omega^{\gamma}\right) \,, \quad \gamma = {\rm min}(3, \alpha) \,,   \quad\,  \quad \alpha \geq 2 \,, \\
\end{cases}
\end{split}
\end{align}
and
\begin{align}
\begin{split}
\label{as2}
\tilde{F}_s\left(\alpha_x,\omega\right) = 
\begin{cases}
\left(c_1 \sin(\pi \alpha/2)/\omega^{\alpha}\right) \left(1 + O\left(1/\omega^{\alpha + 1}\right)\right),    \quad   \quad
\alpha < 2 \,,  \\
c_3/\omega^2 +O\left(1/\omega^{\gamma}\right) \,, \quad \gamma = {\rm min}(3, \alpha)  \,, \quad  \quad \alpha \geq 2 \,, \\
\end{cases}
\end{split}
\end{align}
where
\begin{align}
c_1 = \frac{\Gamma(1 - \alpha) \Gamma(1/2+\alpha)}{\pi} \frac{(1-u)^{\alpha}}{(2 z)^{1-\alpha}}  \,.
\end{align}
The constants $c_2$ and $c_3$ can be also readily calculated but the resulting expressions are a bit lengthy and we do not present them in an explicit form. 

Inspecting the asymptotic forms \eqref{as1} and \eqref{as2} of the PDF $P(\mathcal{S})$ 
we conclude that if $\alpha_x + \alpha_y > 1$, both functions in eqs. \eqref{kern} are integrable in the limit $\omega \to \infty$. Therefore,  for such values of $\alpha_x$ and $\alpha_y$ the PDF $P(\mathcal{S})$ is \textit{finite}
at $\mathcal{S} = \pm \Omega$ and $\mathcal{S} = \pm \Sigma$.

In the borderline case $\alpha_x + \alpha_y = 1$, the situation is a bit more delicate. 
Regardless of the actual values of $\alpha_x $ and $\alpha_y$ which sum to unity, 
both terms in eqs. \eqref{kern} are equal to each other in the leading order. 
This implies that for the sign "-" they cancel each other and the sub-dominant terms come into play. As a result,
the function with the sign "-" is integrable in the limit $\omega \to \infty$ and hence, 
$P(\mathcal{S} = \pm \Sigma)$ is finite. On the contrary, the function with the sign "+"  
decays as $1/\omega$ and therefore the integral diverges. This signifies that 
$P(\mathcal{S})$ diverges logarithmically when $\mathcal{S}$ approaches $\pm \Omega$, but stays
 finite at the edges of the support.

Lastly, for $\alpha_x + \alpha_y < 1$  both functions in eqs. \eqref{kern} decay as 
$c/\omega^{\alpha_x + \alpha_y}$ (with a $\omega$-independent amplitude $c$) which means that the integrals in the second and the fourth lines in eqs. \eqref{om} diverge. As a consequence,  in this case $P(\mathcal{S})$ 
has a  power-law 
divergence at  $\mathcal{S} = \pm \Omega, \pm \Sigma$.

\section{The probability density function $P(\mathcal{S})$: Particular cases}
\label{part}

Here we consider all possible cases A to F listed in the previous section using particular values of $\alpha_x$ and $\alpha_y$ for which the integral in eq. \eqref{pU} can be performed exactly. This will permit us to highlight the richness of the behavior of the PDF $P(\mathcal{S})$, as confronted to the behavior observed in standard BG model. 

\subsection{Case A: $\alpha_x$ and $\alpha_y$ are both less than $1$} 

As an illustrative example of the behavior in the case A, we first consider the choice $\alpha_x = \alpha_y = 1/2$, for which the expression \eqref{pU} attains the form 
\begin{align}
\begin{split}
\label{zu}
P(\mathcal{S}) &= \frac{1}{\pi} \int^{\infty}_0 d\omega \, J_0\left(\frac{v_x}{1 - u} \omega\right) \, J_0\left(\frac{v_y}{1 - u} \omega\right) \, \cos\left(\omega \mathcal{S}\right)  \,.
\end{split}
\end{align}
The integral in the above expression can be performed exactly to give
\begin{align}
\label{casea}
P(\mathcal{S}) = \frac{(1 - u) \,_2F_1\left(1/4, 3/4; 1; \eta\right)}{\pi \sqrt{(v_x + v_y)^2 + 4 v_x v_y - (1-u)^2 \mathcal{S}^2}}
\begin{cases}
1,    \quad  \quad  |\mathcal{S}| \leq \Sigma,\\
0,   \quad  \quad  |\mathcal{S}| > \Sigma, \\
\end{cases}
\end{align}
where the parameter $\Sigma$ defines the edge of the support of $P(\mathcal{S})$ (see eq. \eqref{support}), while
the argument $\eta$ is
\begin{align}
\eta= \frac{16 v_x v_y \left((v_x + v_y)^2 - (1 - u)^2 \mathcal{S}^2\right)}{ \left((v_x + v_y)^2 + 4 v_x v_y- (1 - u)^2 \mathcal{S}^2\right)^2} \,.
\end{align}
Note that the Gauss' hypergeometric function in eq. \eqref{casea} can be expressed through the 
complete elliptic integral of the first kind. 

\begin{figure}[h!]
\centering
\includegraphics[width=140mm]{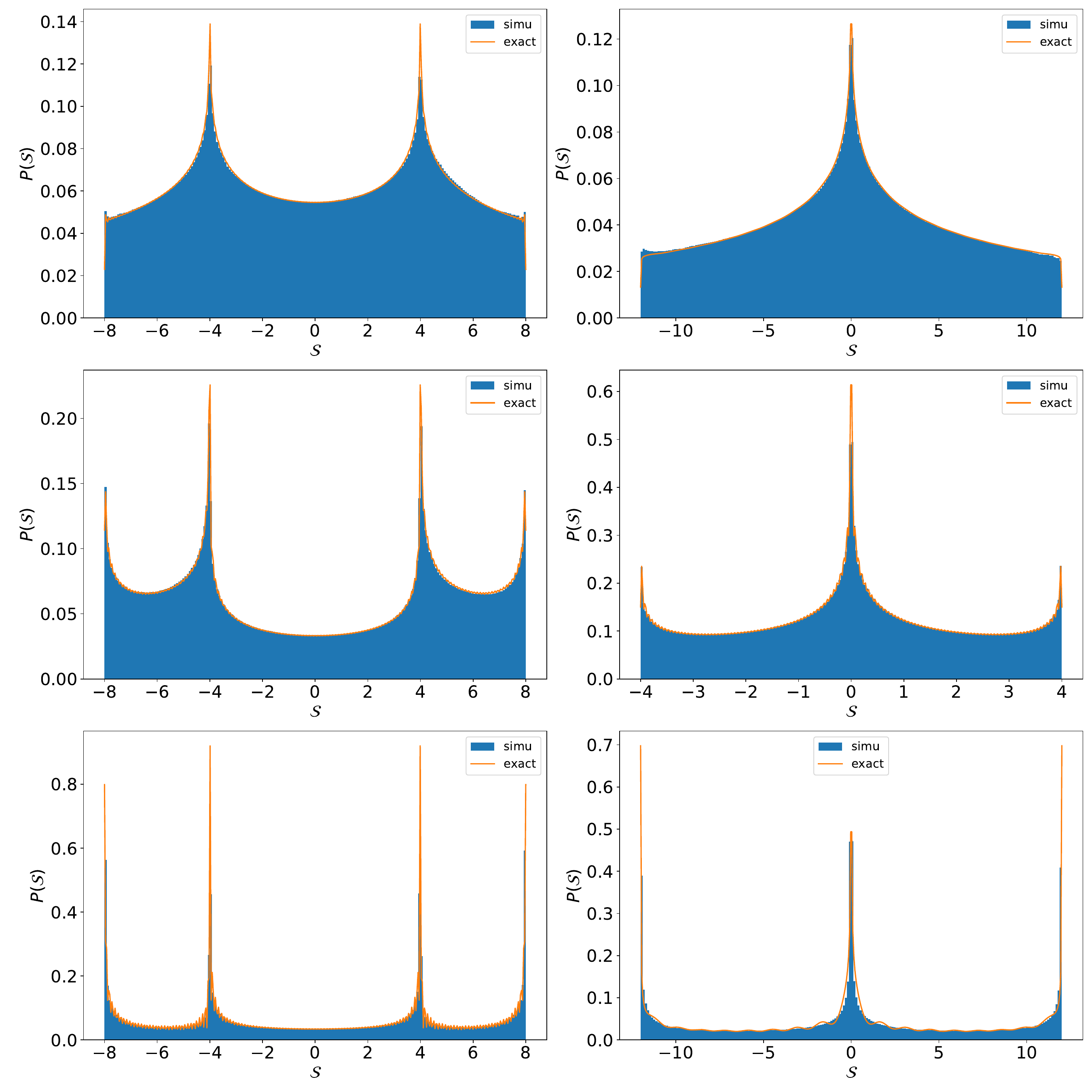}
\caption{Case A: The probability density function $P(\mathcal{S})$  
as function of $\mathcal{S}$ for $u=1/2$. The magenta curve show our analytical predictions in eq. \eqref{pU}, 
while the histogram (dark blue) here and henceforth 
depicts the results of numerical simulations of the process in Eqs. \eqref{LE}.
Top row: $P(\mathcal{S})$ for  $\alpha_x  = \alpha_y = 1/2$ (eq. \eqref{casea}, magenta curve), $v_x = 1$ and $v_y = 3$ (left panel). $P(\mathcal{S})$ in eq. \eqref{casea} (magenta curve) for $v_x= v_y =3$ (right panel).  
Central row: Numerical evaluation (magenta curve) of the integral in eq. \eqref{pU} for $\alpha_x = 1/4$ and $\alpha_y = 1/2$, $v_x = 3$ and $v_y = 1$   (left panel) and $v_x = v_y = 1$ (right panel). Bottom row: Numerical evaluation (magenta curve) of the integral in eq. \eqref{pU} for $\alpha_x = 1/8$ and $\alpha_y = 1/4$, $v_x = 1$ and $v_y = 3$ (left panel) and $v_x = v_y = 3$ (right panel). Note that all the peaks in this figure have an infinite height.}
\label{fig:caseA} 
\end{figure}

 In Fig. \ref{fig:caseA} (top row) we depict the PDF in eq. \eqref{casea} with the coupling parameter
 $u=1/2$ (which is the same for all panels in this Figure). In the left panel we plot $P(\mathcal{S})$ 
for the case when $v_x$ and $v_y$ are not equal; here, our choice is $v_x = 1$ and $v_y = 3$.
We observe that $P(\mathcal{S})$ is a bimodal function with the minimum at $\mathcal{S}= 0$ and two peaks at $\mathcal{S} = \pm \Omega$ (see eq. \eqref{Omega} for the definition of $\Omega$). At these two points the probability density function diverges logarithmically when $|\mathcal{S}|$ approaches $\Omega$ from above or from below (see Subsec. \ref{edges}).
Next, when  $\mathcal{S}$ approaches the edges of the support, i.e., $\mathcal{S} \to \pm \Sigma$, (here $\Sigma = 8$), the PDF $P(\mathcal{S})$ tends to a constant value 
\begin{align}
\label{peak}
\lim_{\mathcal{S} \to \pm \Sigma} P(\mathcal{S}) = \frac{(1 - u)}{2 \pi \sqrt{v_x v_y}} \,,
\end{align}
and then abruptly drops to zero. In the right panel to Fig. \ref{fig:caseA} we consider the behavior of $P(\mathcal{S})$ for equal amplitudes of the noises, $v_x = v_y = 3$.  In this case $\Omega \equiv  0$ such that the two peaks present in the left panel merge into a single symmetric peak located at the origin. 
Note that $P(\mathcal{S})$ also diverges logarithmically when $\mathcal{S} \to 0$ and approaches a constant value defined in eq. \eqref{peak} when $\mathcal{S} \to \pm \Sigma$ (here, $\Sigma = 12$). 

In the central row in Fig. \ref{fig:caseA} we depict  $P(\mathcal{S})$ for a somewhat lower value of the switching rate of the noise acting on the $x$-component. Here, we take $\alpha_x = 1/4$, while $\alpha_y$ is kept equal to the same value as in the top row; that being, $\alpha_y = 1/2$.   In the left panel we present the PDF in eq. \eqref{pU} with the integral being evaluated in Scipy\cite{Virtanen2020} (magenta curve) for $v_x =3$ and  $v_y = 1$. We observe that the PDF is four-modal with four infinitely high peaks at $\mathcal{S} = \pm \Omega$ and $\mathcal{S} = \pm \Sigma$,
and has three minima. As remarked above in the Subsec. \ref{edges}, $P(\mathcal{S})$ diverges as a power-law when $\mathcal{S}$ approaches the points of irregularity $\pm \Omega$ and also the edges of the support. In the right panel we depict $P(\mathcal{S})$ for the same values of $\alpha_x$ and $\alpha_y$ but  take  
the amplitudes of noises to be equal to each other, $v_x = v_y = 1$. In this case $\Omega = 0$, the two peaks in the central part present in the left panel merge, such that we have a single peak at $\mathcal{S} = 0$. Note that $P(\mathcal{S})$ diverges as a power-law when $\mathcal{S} \to 0$. A salient feature here is that $P(\mathcal{S})$ diverges, also as a power-law, when $\mathcal{S}$ tends to the edge of the support, i.e., $\mathcal{S} \to \pm \Sigma$ (here, $\Sigma =4$), instead of approaching a constant value - the behavior we observed in the top row. Consequently, $P(\mathcal{S})$ is a three-modal function with two minima. We explained such a divergence at the edges of the support in Subsec. \ref{edges}.

In the bottom row in Fig. \ref{fig:caseA} we further diminish the values of both switching rates choosing 
$\alpha_x = 1/8$ and $\alpha_y = 1/4$. Here, the amplitudes are taken the same as above:  $v_x = 1$ and $v_y = 3$ (left panel) and  $v_x = v_y = 3$ (right panel).
We observe that the behavior of $P(\mathcal{S})$ remains qualitatively the same as in the central row, 
but peaks become essentially more pronounced and the minima between the peaks become somewhat smaller: Indeed, as remarked above, the power-law divergences in this case become stronger.

\subsection{Case B: $\alpha_x < 1$ and $\alpha_y = 1$, or vice versa}

To illustrate the behavior in this case 
we take $\alpha_x = 1/2$ and $\alpha_y = 1$. For these values of parameters the probability density function in eq. \eqref{pU} is
\begin{align}
\begin{split}
\label{caseb}
P(\mathcal{S}) &= \frac{(1-u)}{2 \pi v_y} \int^{\infty}_0 \frac{d\omega}{\omega} \, J_0\left(\frac{v_x}{1-u} \omega\right) \\&\times \left[\sin\left(\left(\frac{v_y}{1-u} - \mathcal{S}\right) \omega\right) + \sin\left(\left(\frac{v_y}{1-u} + \mathcal{S}\right) \omega\right)
\right] \,.
\end{split}
\end{align} 
Performing the integral in the latter equation we realize that we have to distinguish between the situations in which $v_x$ is smaller than $v_y$, or conversely, is greater than $v_v$. We focus first on the situation when $v_x \leq v_y$. We find then
that $P(\mathcal{S})$ is 
\begin{align}
\label{caseb1}
P(\mathcal{S}) = \frac{1 - u}{2 \pi v_y}
\begin{cases}
\pi , \quad \qquad \quad \qquad \qquad \qquad \qquad  \quad   - \Omega \leq \mathcal{S} \leq \Omega,\\
\frac{\pi}{2} + \arcsin\left(\frac{v_y - (1 - u) \mathcal{S}}{v_x}\right),  \qquad \qquad  \quad  \Omega \leq \mathcal{S} \leq \Sigma,  \\
\frac{\pi}{2} + \arcsin\left(\frac{v_y + (1 - u) \mathcal{S}}{v_x}\right),  \quad \qquad \quad \,\, - \Sigma \leq \mathcal{S} \leq - \Omega \,, \\
0 , \quad \qquad \quad \qquad \qquad \qquad \qquad \quad \quad \,\,\,\,\,  \quad |\mathcal{S}| \geq \Sigma \,.
\end{cases}
\end{align}
Next, for $v_x \geq v_y$ we find
 \begin{align}
\label{caseb2}
P(\mathcal{S}) = \frac{1 - u}{2 \pi v_y}
\begin{cases}
\arcsin\left(\frac{v_y - (1 - u) \mathcal{S}}{v_x}\right) +  \arcsin\left(\frac{v_y + (1 - u) \mathcal{S}}{v_x}\right), \quad \quad  - \Omega \leq \mathcal{S} \leq \Omega,\\
\frac{\pi}{2} + \arcsin\left(\frac{v_y - (1 - u) \mathcal{S}}{v_x}\right),  \qquad \qquad \qquad  \quad \quad \,\,\, \quad \Omega \leq \mathcal{S} \leq \Sigma,  \\
\frac{\pi}{2} + \arcsin\left(\frac{v_y + (1 - u) \mathcal{S}}{v_x}\right),  \qquad \quad \,\,\,\, \qquad \quad \qquad  - \Sigma \leq \mathcal{S} \leq - \Omega , \\
0 , \qquad \qquad \quad \qquad \qquad \quad \quad \qquad \qquad \qquad \qquad \quad  |\mathcal{S}| \geq \Sigma \,.
\end{cases}
\end{align}

\begin{figure}[h!]
\centering
\includegraphics[width=150mm]{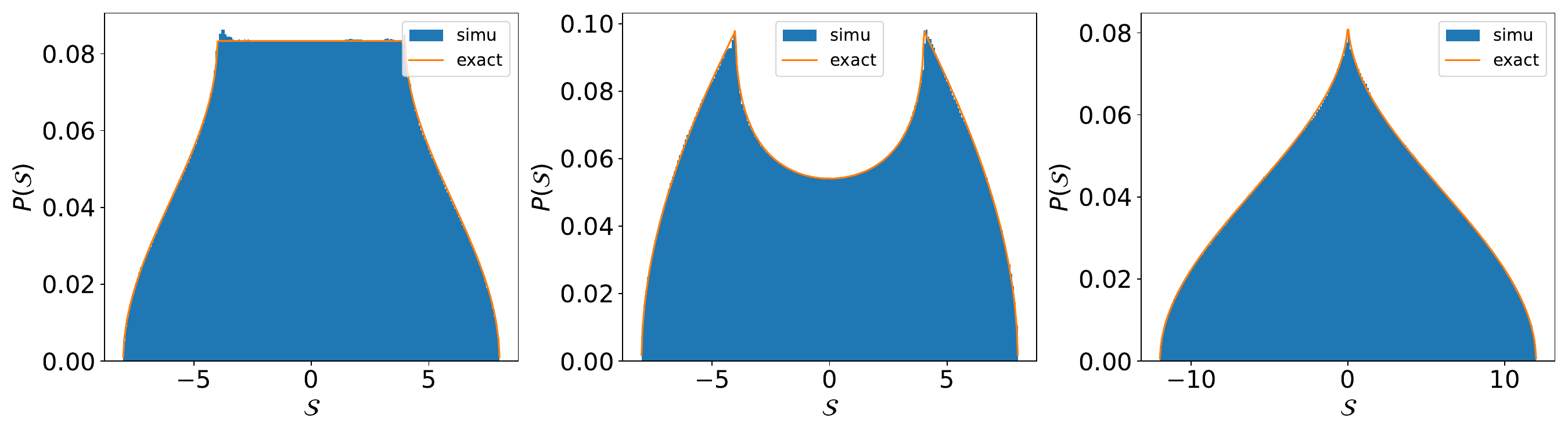}
\caption{Case B:  The probability density function $P(\mathcal{S})$ in eqs. \eqref{caseb1} and \eqref{caseb2} 
(magenta curves) as function of $\mathcal{S}$ for $u=1/2$, $\alpha_x  = 1/2$ and $\alpha_y=1$. 
Left panel: $v_x=1$ and $v_y=3$. Middle panel: 
$v_x=3$ and $v_y=1$ Right panel: 
$v_x=3$ and $v_y=3$.}
\label{fig:CaseB} 
\end{figure} 

In Fig. \ref{fig:CaseB} we depict the typical behavior in case B as exemplified here by the choice $\alpha_x = 1/2$ and $\alpha_y = 1$. We observe that the functional form of $P(\mathcal{S})$ appears to be quite non-trivial here and depends on the relation between the amplitudes of noises. In particular, for $v_y > v_x$ (the left panel) the PDF exhibits a plateau-like behavior in its central part, which is followed for $|\mathcal{S}| > \Omega$ by a rapid decay to zero,  approached when $\mathcal{S} = \pm \Sigma$. Conversely, for $v_y < v_x$ (the middle panel) the PDF is $U$-shaped in its central part with a minimum at $\mathcal{S} = 0$, has two asymmetric, cusp-like peaks at $\mathcal{S} = \pm \Omega$, followed by a rapid decrease to zero approached when $\mathcal{S}= \pm \Sigma$. 
This illustrates the statement made above that for unequal values of $\alpha_x$ and $\alpha_y$ the PDF $P(\mathcal{S})$ may have quite different  shapes for situations when $v_x > v_y$ or $v_x < v_y$. In the right panel we take  $v_x = v_y$.
In this case, two peaks present in the middle panel merge 
and one is left with a single cusp-like peak at $\mathcal{S} = 0$.
Note, as well,  that the functional forms of $P(\mathcal{S})$ presented in Fig. \ref{fig:CaseB} are quite generic for the case B, regardless of the actual choice of $\alpha_x$ provided that $\alpha_x < \alpha_y = 1$.

\subsection{Case C: $\alpha_x  < 1$ and $\alpha_y > 1$, or vice versa}

Here we choose as a representative particular case the values $\alpha_x = 1/2$ and $\alpha_y = 2$. Performing the integral in eq. \eqref{pU} for $v_y>v_x$ we find 
\begin{align}
	\label{casec1}
P(\mathcal{S}) = \frac{(1-u)}{8 v_y^3}
\begin{cases}
3 \left(v_x^2 + 2v_y^2 - 2 \mathcal{S}^2\right) \,, \,\,	\quad   - \Omega \leq \mathcal{S} \leq \Omega,\\
\Psi_{C,1} \,,	 \qquad   \qquad \qquad  \qquad  \Omega \leq \mathcal{S} \leq \Sigma ,  \\
\Psi_{C,2} \,,  \qquad   \qquad  \qquad \,\,  \quad  - \Sigma \leq \mathcal{S} \leq - \Omega , \\
	0 ,  \qquad \qquad   \qquad \qquad   \,\, \qquad \quad |\mathcal{S}| \geq \Sigma \,.
\end{cases}
\end{align}
where 
\begin{align}
\begin{split}
\Psi_{C,1} &= \frac{1}{\pi}\left(3(3 \mathcal{S}+v_y)\sqrt{v_x^2-(\mathcal{S}-v_y)^2}-(2\mathcal{S}^2+v_x^2-2v_y^2)\arccos{\left[\frac{\mathcal{S}-v_y}{v_x}\right]}\right) \,, \\
\Psi_{C,2} &= \frac{1}{\pi}\left(3(v_y -3 \mathcal{S})\sqrt{v_x^2-(\mathcal{S}+v_y)^2}-(2 \mathcal{S}^2+v_x^2-2v_y^2)\arccos{\left[\frac{|\mathcal{S}+v_y|}{v_x}\right]}\right) \,,
\end{split}
\end{align}
whereas for $v_x>v_y$ we get
\begin{align}
	\label{casec2}
	P(\mathcal{S}) = \frac{(1-u)}{8\pi v_y^3}
	\begin{cases}
		\Psi_{C,3}	 \,,        \qquad  \quad  \,\, - \Omega \leq \mathcal{S} \leq \Omega \,,
		\\
	    \Psi_{C,4} \,,	      \qquad  \qquad  \Omega \leq \mathcal{S} \leq \Sigma , 
	     \\
		\Psi_{C,5} \,,         \qquad  \quad \,\,   - \Sigma \leq \mathcal{S} \leq - \Omega , \\
		0 , \quad  \quad  \qquad \qquad  \quad \,\, |\mathcal{S}| \geq \Sigma \,,
	\end{cases}
\end{align}
with 
\begin{align}
\begin{split}
\Psi_{C,3} & = 3 \Bigg(3 \mathcal{S} \left(\sqrt{v^2_x-(\mathcal{S}-v_y)^2}-\sqrt{v^2_x-(\mathcal{S}+v_y)^2}\right)\\ &+v_y \left(\sqrt{v^2_x-(\mathcal{S}-v_y)^2}+\sqrt{v^2_x-(\mathcal{S}+v_y)^2}\right)-(2\mathcal{S}^2+v_x^2-2v_y^2) \\ &\times \left(\arcsin\left[\frac{\mathcal{S}-v_y}{v_x}\right]+\arcsin\left[\frac{\mathcal{S}+v_y}{v_x}\right]\right) \Bigg) \,,\\
\Psi_{C,4} & = \left(3(3 \mathcal{S}+v_y)\sqrt{v_x^2-(\mathcal{S}-v_y)^2}-(2\mathcal{S}^2+v_x^2-2v_y^2)\arccos{\left[\frac{\mathcal{S}-v_y}{v_x}\right]}\right) \,,\\
\Psi_{C,5} & =\left(3(v_y-3\mathcal{S})\sqrt{v_x^2-(\mathcal{S}-v_y)^2}-(2\mathcal{S}^2+v_x^2-2v_y^2)\arccos{\left[\frac{|\mathcal{S}+v_y|}{v_x}\right]}\right) \,.  
\end{split}
\end{align}

\begin{figure}[h!]
	\centering
	\includegraphics[width=150mm]{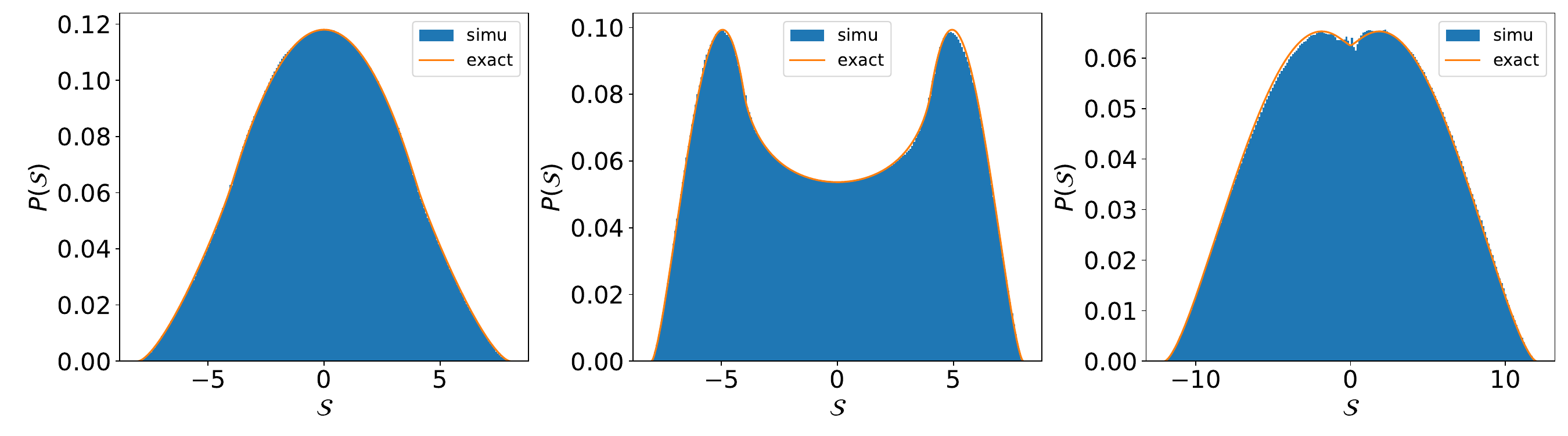} 	
	\caption{Case C: The probability density function $P(\mathcal{S})$ in eqs. \eqref{casec1} and \eqref{casec2}  
		as function of $\mathcal{S}$ for $u=1/2$, $\alpha_x  = 1/2$ and $\alpha_y=2$. 
		Left panel: $v_x=1$ and $v_y=3$. Middle panel:  $v_x=3$ and $v_y=1$, Right panel:  $v_x=3$ and $v_y=3$.}
	\label{fig:CaseC} 
\end{figure} 
In Fig. \ref{fig:CaseC} we plot the PDF $P(\mathcal{S})$ as function of $\mathcal{S}$ for the Case C, as exemplified here by the choice $\alpha_x = 1/2$ and $\alpha_y = 2$. The magenta curves are the functions defined in eqs. \eqref{casec1} and \eqref{casec2} for $v_x=1$ and $v_y=3$ (left panel), for $v_x=3$ and $v_y=1$ (middle panel) and for $v_x = v_y = 3$ (right panel), respectively. 
We observe that the piece-wise continuous function $P(\mathcal{S})$ in the left panel is smooth and 
 has a standard bell-like shape. Moreover, the region in the vicinity of the origin can be well-approximated by a Gaussian function in eq. \eqref{Gauss}. The only consequence of the fact that the system is driven by dichotomous noises is that 
 $P(\mathcal{S})$ has a finite support (here, $\Sigma = 8$).   Interchanging the values of the amplitudes of noises, we  encounter a completely different functional form of $P(\mathcal{S})$ (see the middle panel): it becomes $M$-shaped with two rounded peaks located at  $\mathcal{S} = \pm  \Omega$  and a minimum at $\mathcal{S} = 0$. In the right panel we take the amplitudes of noises to be equal. Interestingly enough, here one again observes a different shape of the PDF - it is bimodal with a small deep at the origin. Such a behavior persists for any $\alpha_x<1$.

 \subsection{Case D:  $\alpha_x = \alpha_y = 1$} 
 
\begin{figure}[ht]
\centering
\includegraphics[width=120mm]{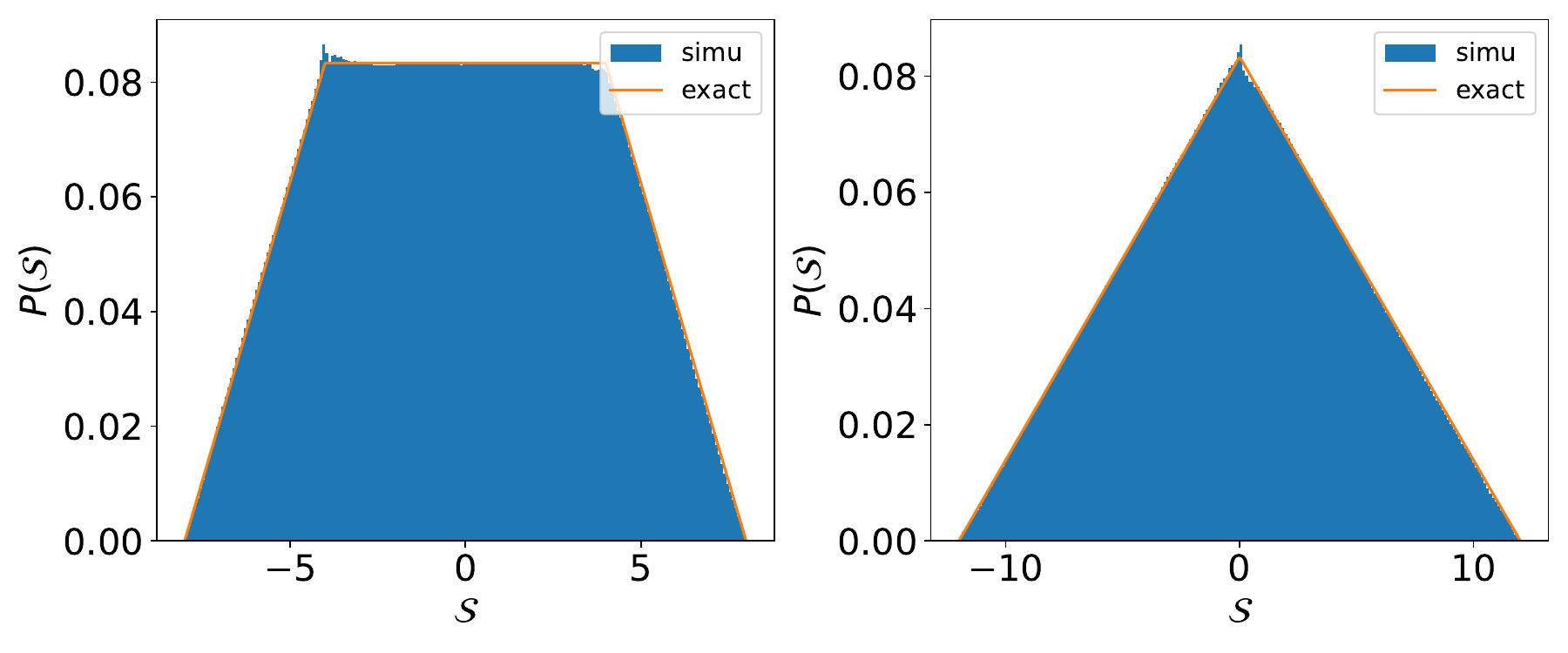} 
\caption{Case D: The probability density function $P(\mathcal{S})$ in eq. \eqref{cased}  as function of $\mathcal{S}$ for $u=1/2$ and $\alpha_x  = \alpha_y= 1$, In the left panel we take $v_x=1$ and $v_y=3$, while in the right panel the amplitudes of noises are taken equal, $v_x= v_y=3$.  }
\label{fig:CaseD} 
\end{figure}

 Setting $\alpha_x = \alpha_y = 1$ in eq. \eqref{pU} we get
 \begin{align}
 \begin{split}
 P(\mathcal{S}) &= \frac{(1 - u)^2}{\pi v_x v_y} \int^{\infty}_0 \frac{d \omega}{\omega^2} \sin\left(\frac{v_x}{1 - u} \omega\right) \,\sin\left(\frac{v_y}{1 - u} \omega\right) \, \cos\left(\omega \mathcal{S}\right) \\
 &= \frac{(1-u)}{8 v_x v_y} \Big(|v_x + v_y -(1-u) \mathcal{S}| + |v_x + v_y + (1-u) \mathcal{S}| \\
 & - |(1 - u) \mathcal{S} - v_y + v_x| - |(1 - u) \mathcal{S} + v_y - v_x| \Big) \,.
 \end{split}
 \end{align}
 Because the switching rates are the same, it suffice to consider just the case $v_y \geq v_x$
 in which $P(\mathcal{S})$ attains a piecewise continuous trapezoidal form 
  \begin{align}
\label{cased}
P(\mathcal{S}) = \frac{1 - u}{4 v_x v_y}
\begin{cases}
2 v_x, \qquad  \qquad\qquad\qquad \quad \quad  |\mathcal{S}| \leq \Omega,\\
v_x + v_y - (1 - u) \mathcal{S},  \quad  \quad  \Omega \leq \mathcal{S} \leq \Sigma ,  \\
v_x + v_y + (1 - u) \mathcal{S},  \quad \,\, - \Sigma \leq \mathcal{S} \leq - \Omega, \\
0 , \qquad \qquad \qquad \qquad \qquad   \,\,\,\,\,\, |\mathcal{S}| \geq \Sigma \,.
\end{cases}
\end{align}
This function is depicted in Fig. \ref{fig:CaseD} in the left panel for $v_x=1$ and $v_y=3$. In the right panel we plot $P(\mathcal{S})$ for equal amplitudes of noise $v_x=v_y=3$. In this case two later the central region disappears, such that the PDF attains a triangular shape, with a cusp at the origin. 

\subsection{Case E:  $\alpha_x = 1$ and $\alpha_y > 1$,  or vice versa}

To highlight the behavior of $P(\mathcal{S})$ in this case we choose $\alpha_x = 1$ and $\alpha_y = 3/2$ in which particular instance the expression
\eqref{pU} reads
\begin{align}
P(\mathcal{S}) = \frac{2 (1-u)^2}{\pi v_x v_y} \int^{\infty}_0 \frac{d\omega}{\omega^2} \, J_1\left(\frac{v_y}{1-u} \omega\right) \, \sin\left(\frac{v_x}{1-u} \omega\right) \, \cos\left(\omega \mathcal{S}\right) \,.
\end{align}
Performing the integral, we get, for $v_x \geq v_y$,
  \begin{align}
\label{casee}
P(\mathcal{S}) = \frac{1 - u}{4 v_x}
\begin{cases}
2 \,, \qquad  \quad \qquad \quad \quad \qquad |\mathcal{S}| \leq \Omega,\\
1 +  \Psi_{E,+} \,, \,\, \quad  \qquad  - \Sigma \leq \mathcal{S} \leq -\Omega,  \\
1 + \Psi_{E,-} \,,    \qquad  \qquad \Omega \leq \mathcal{S} \leq \Sigma, \\
0 , \qquad \qquad \qquad \qquad  \quad |\mathcal{S}| \geq \Sigma \,,
\end{cases}
\end{align}
where
\begin{align}
\Psi_{E,\pm} = \frac{4}{\pi v_y} \left(v_x \pm (1 - u)\mathcal{S}\right) \,_2F_1\left(1/2, - 1/2; 3/2; \frac{\left(v_x \pm (1 - u)\mathcal{S}\right)^2}{v_y^2}\right) \,.
\end{align}
Note that the hypergeometric function in the above equation can be expressed via elementary and the inverse trigonometric functions.

Respectively, 
for $v_y \geq v_x$ we obtain
 \begin{align}
\label{casee2}
P(\mathcal{S}) = \frac{1 - u}{4 v_x}
\begin{cases}
\Psi_{E,+} + \Psi_{E,-}  \,,   \quad   \quad \qquad |\mathcal{S}| \leq \Omega,\\
1 +  \Psi_{E,+} \,, \,\, \quad  \qquad  - \Sigma \leq \mathcal{S} \leq -\Omega,  \\
1 + \Psi_{E,-} \,,    \qquad  \qquad \Omega \leq \mathcal{S} \leq \Sigma, \\
0 , \qquad \qquad \qquad \qquad  \quad |\mathcal{S}| \geq \Sigma \,.
\end{cases}
\end{align}

\begin{figure}[h!]
	\centering
	\includegraphics[width=150mm]{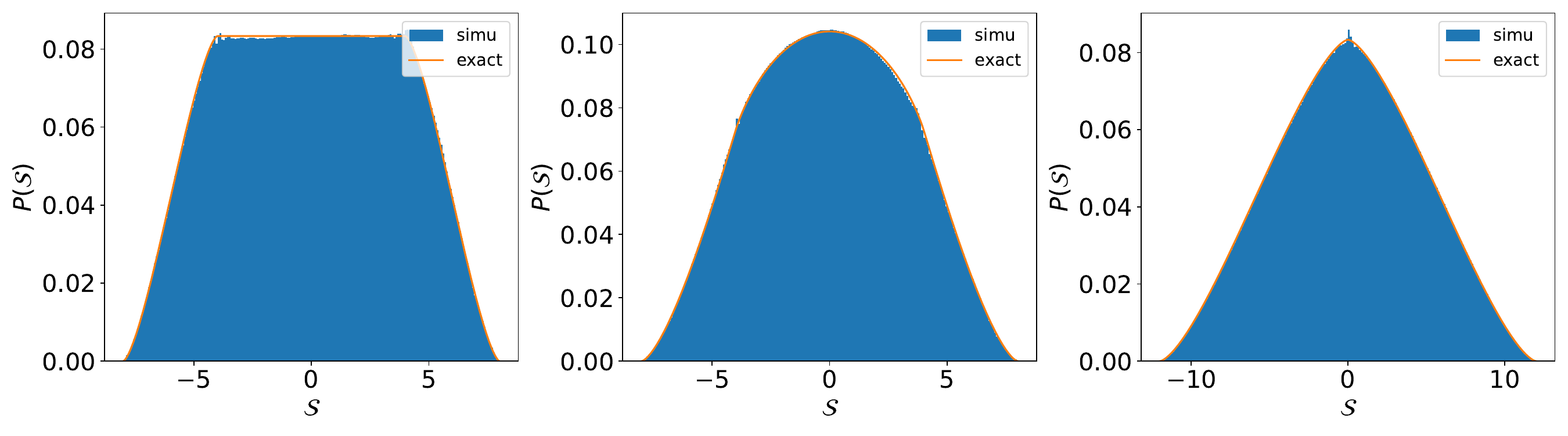} 
	\caption{Case E: The probability density function $P(\mathcal{S})$ in eqs. \eqref{casee} in \eqref{casee2} as function of $\mathcal{S}$ for $u=1/2$, $\alpha_x = 1$ and $\alpha_y = 3/2$.  Left panel:  $v_x=3$ and $v_y=1$, eq. \eqref{casee}. Middle panel:  $v_x=1$ and $v_y=3$, Right panel: 
			$v_x=3$ and $v_y=3$, eq. \eqref{casee2}.}
	\label{fig:CaseE} 
\end{figure}

In Fig. \ref{fig:CaseE} we depict $P(\mathcal{S})$ in the case E. We observe that the shapes of the PDF in the left and the right panels appear to be very similar to those predicted for the Case D, 
except that the cusp-like corners become quite rounded. In the middle panel $P(\mathcal{S})$ is a bell-shaped function centered at $\mathcal{S} = 0$.

 \subsection{Case F: $\alpha_x > 1$ and $\alpha_y > 1$}
 
Finally, we turn to  the case F, for which we consider two choices of values of  $\alpha_x$ and  $\alpha_y=2$ :
$\alpha_x  =  3/2$ and $ \alpha_y=2$, and 
$\alpha_x  = \alpha_y=2$. The integral in eq. \eqref{pU} can be performed exactly in both cases, but the expression 
for the former case appears to be very lengthy; we do not list it here in an explicit form and just use for plots and for the comparison with the numerical results.  For the latter case $\alpha_x  = \alpha_y=2$ the PDF $P(\mathcal{S})$ has a more compact form and is given explicitly by
  \begin{align}
	\label{casef1}
	P(\mathcal{S}) = \frac{1 - u}{v_y^3}
	\begin{cases}
3(5v_y^2- v_x^2 - 5(1-u)^2 \mathcal{S}^2) \,,	   \qquad   |\mathcal{S}| \leq \Omega,\\
\Psi_{F,1}	\,,  	\quad \,\, \qquad  \qquad \qquad    \qquad  \Omega \leq \mathcal{S} \leq \Sigma ,  \\
\Psi_{F,2}	\,,	 \qquad \quad \,\,\,\, \qquad \qquad     \quad	 	   - \Sigma \leq \mathcal{S} \leq - \Omega, \\
		0 \,, \qquad \,\, \qquad \qquad \qquad \qquad  \qquad |\mathcal{S}| \geq \Sigma \,,
	\end{cases}
\end{align}
with 
\begin{align}
\begin{split}
\Psi_{F,1} &= \frac{3(v_x + v_y - (1-u) \mathcal{S})^3}{8v_x^3}\Big((1-u)^2 \mathcal{S}^2+3(1-u)(v_x+v_y) \mathcal{S} \\&- 4(v_x^2-4v_x v_y +v_y^2)\Big) \,,\\
\Psi_{F,2} &= \frac{3((1-u) \mathcal{S}+v_x+v_y)^3}{8v_x^3}\Big((1-u)^2 \mathcal{S}^2-3(1-u) (v_x+v_y) \mathcal{S}\\&-4(v_x^2-4v_x v_y +v_y^2)\Big) \,.
\end{split}
\end{align}

\begin{figure}[h!]
	\centering
	\includegraphics[width=150mm]{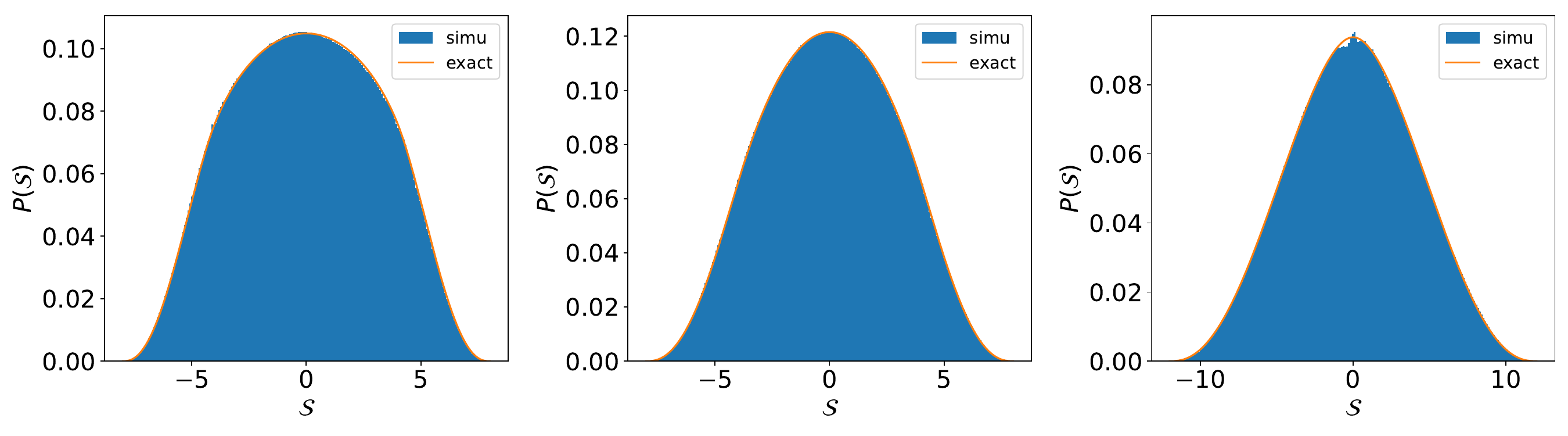}
	\caption{Case F: The probability density function $P(\mathcal{S})$  as function of $\mathcal{S}$ for $u=1/2$, $\alpha_x =3/2$ and  $\alpha_y=2$. Left panel: $v_x=3$ and $v_y=1$.
		Middle panel: $v_x=1$ and $v_y=3$. Right panel: $v_x=3$ and $v_y=3.$ 
	}
	\label{fig:CaseF} 
\end{figure} 

We depict $P(\mathcal{S})$ evaluated for $\alpha_x  =  3/2$ and $ \alpha_y=2$ and three choices of the amplitudes $v_x$ and $v_y$
in Fig. \ref{fig:CaseF}. In all the cases, the PDF $P(\mathcal{S})$ appears to be a bell-shaped, piece-wise continuous function with a finite support. The very central part of $P(\mathcal{S})$ is well-approximated by a Gaussian function in eq. \eqref{pU}. Note, however, that the forms depicted in the left and the middle panels,  
which are obtained by a mere interchange $v_x\leftrightarrow v_y$, are visibly different, with the PDF in the left panel being apparently broader than the one in the middle panel.

\section{The behavior in case when $\zeta_x(t)$ in eqs. \eqref{LE} is not fluctuating.}
\label{Zx0}

In this Section we examine the behavior in case when either of the noises $\zeta(t)$ in the right-hand-side of eqs. \eqref{LE}, say, $\zeta_x(t)$ is not fluctuating in time, i.e., is not a noise. We will distinguish between two situations: the one in which the switching rate $\lambda_x = 0$ and the one in which the amplitude $v_x = 0$. Note that in both situations the $x$-component will be entrained into a stochastic dynamics through the coupling to the $y$-component which experiences the action of the noise $\zeta_y(t)$. Note, as well, that in this case the evolution in the standard BG model is also quite peculiar, as discussed in \cite{Cerasoli2021a}. It appears to be even more fascinating in the case of dichotomous noise, as we  show below. 

\subsection{Vanishing switching rate $\lambda_x$}

We first focus on the situation when $\lambda_x = 0$ (i.e., $\alpha_x = 0$). Here, 
$\zeta_x(t) \equiv \zeta_x(0)$ and equals either $+ v_x $ or $- v_x $, with equal probability, such that the $x$-component is subject to a time-independent randomly orientated (positive or negative) force.  
The noise $\zeta_y(t)$ is still characterized by a non-zero switching rate, $\lambda_y > 0$. In Fig. \ref{fig:01} we have already depicted the behavior of $P(\mathcal{S}=0)$ and realized that it is markedly different from the behavior for $\alpha_x > 0$.

In this particular case the integral  in eq. \eqref{pU} can be performed exactly and the probability density function reads
\begin{align}
\begin{split}
\label{pUalphax0}
P(\mathcal{S}) &= \frac{\Gamma(1/2 + \alpha_y)}{\pi} \left(\frac{v_y}{2 (1 - u)}\right)^{1/2-\alpha_y}  
 \int^{\infty}_0 d\omega \,  \frac{J_{\alpha_y - 1/2}\left(\frac{v_y}{1 - u} \omega\right)}{\omega^{\alpha_y - 1/2}} \, \cos\left(\frac{v_x}{1-u} \omega\right) \, \cos(\omega \mathcal{S})  \\
 & = \frac{\Gamma(1/2+\alpha_y) (1- u)}{2 \sqrt{\pi} \Gamma(\alpha_y) v_y^{2 \alpha_y-1}}
  \begin{cases}
\left(v_y^2 - \left(v_x + (1 - u) \mathcal{S}\right)^2\right)^{\alpha_y - 1}, \quad v_y^2 \geq \left(v_x + (1 - u) \mathcal{S}\right)^2,\\
\left(v_y^2 - \left(v_x - (1 - u) \mathcal{S}\right)^2\right)^{\alpha_y - 1}, \quad v_y^2 \geq \left(v_x - (1 - u) \mathcal{S}\right)^2,  \\
\Psi(\mathcal{S}),  \qquad  \qquad \qquad \quad \quad \,\,\,\, \qquad \mathcal{S} \in \mathcal{I}, \\
0 , \qquad \qquad \qquad \qquad \qquad \qquad \text{otherwise} \,.
\end{cases}
\end{split}
\end{align}
where
\begin{align}
\Psi(\mathcal{S}) = \left(v_y^2 - \left(v_x + (1 - u) \mathcal{S}\right)^2\right)^{\alpha_y - 1} + \left(v_y^2 - \left(v_x - (1 - u) \mathcal{S}\right)^2\right)^{\alpha_y - 1} \,,
\end{align}
and $\mathcal{I}$ denotes the set of such values of $\mathcal{S}$ (if any) which fulfill simultaneously the inequalities 
in the first and the second  lines. For  such values of $\mathcal{S}$ the function $\Psi(\mathcal{S})$ replaces the expressions in the first and the second lines. Note that expression \eqref{pUalphax0}  is valid for any values of $v_x$, $v_y$ and $\alpha_y$. Note, as well, that in the limit $\alpha_y \to 0$ the probability density function is equal to the sum of four delta functions. 

\begin{figure}[H]
\centering
\includegraphics[width=150mm]{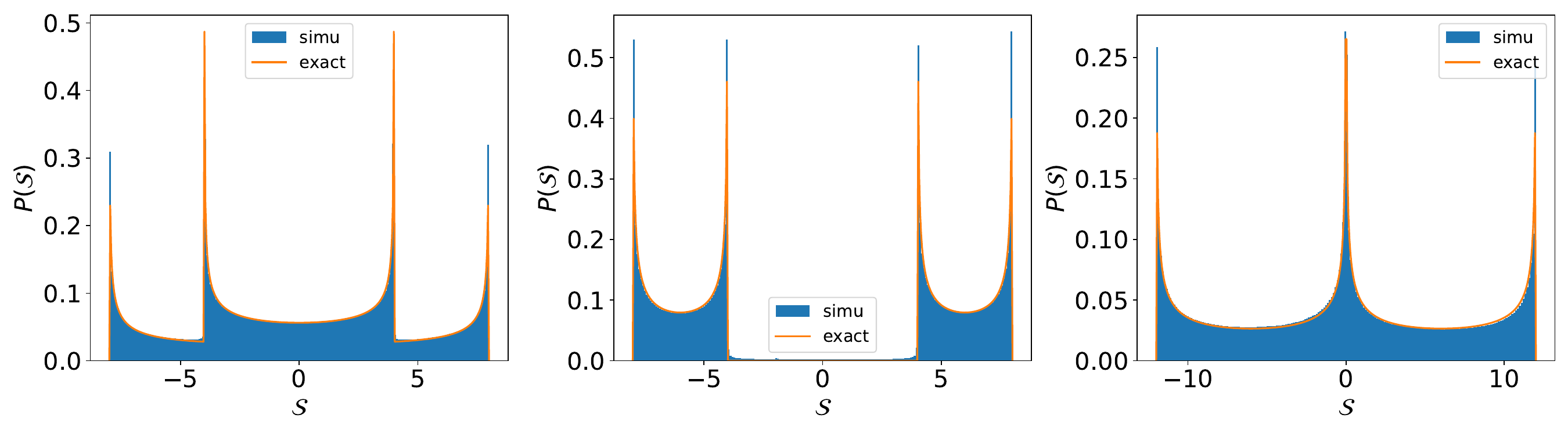} 
\caption{Vanishing switching rate $\lambda_x$. The probability density function $P(\mathcal{S})$ 
as function of $\mathcal{S}$ for $u=1/2$, $\alpha_x = 0$ and $\alpha_y = 1/2$. Left panel: $P(\mathcal{S})$ for  $v_x = 1$ and $v_y = 3$.  Middle panel: $P(\mathcal{S})$ for  $v_x = 3$ and $v_y = 1$. Note that $P(\mathcal{S}) \equiv 0$ for $|\mathcal{S}| < \Omega $. Right panel: $P(\mathcal{S})$ for  $v_x = v_y = 3$.}
\label{fig:12} 
\end{figure} 

Consider first the choice $\alpha_y = 1/2$, such that we deal here with a very particular instance of the case A. Behavior of $P(\mathcal{S})$ in eq. \eqref{pUalphax0} is depicted in Fig. \ref{fig:12}.
We observe very unusual shapes of $P(\mathcal{S})$ 
for both situations when $v_x < v_y$ (left panel) and when $v_x > v_y$ (middle panel). The PDF in the left panel has four infinitely high peaks and a single minimum at $\mathcal{S} = 0$. Interestingly enough, the peaks at $\mathcal{S} = \pm \Omega$ are highly asymmetric in this case: while $P(\mathcal{S}) \to \infty$ when $|\mathcal{S}|$ approaches $\Omega$ from below, the PDF $P(\mathcal{S})$ approaches a constant value when $|\mathcal{S}|$ tends to $\Omega$ from above.  
In the middle panel we observe a very different form: here, the PDF consists of two $U$-shaped curves diverging at $\mathcal{S} = \pm \Omega$ and $\mathcal{S} = \pm \Sigma$ and separated by an interval $\mathcal{S} \in (-\Omega,\Omega)$ in which the PDF  is \textit{exactly} equal to zero.  
Recalling our discussion in the Introduction about the heavy tails of the PDF of the angular velocity, we may expect that
in this precisely the case such tails may be absent because the SG cannot visit the origin. In the right panel we depict the behavior for equal amplitudes of noises, $v_x = v_y =3$. Here, the behavior appears to be very similar to the one which we have already encountered in Fig. \ref{fig:caseA} (see the right panels in the central and in the bottom rows).

\begin{figure}[H]
\centering
\includegraphics[width=150mm]{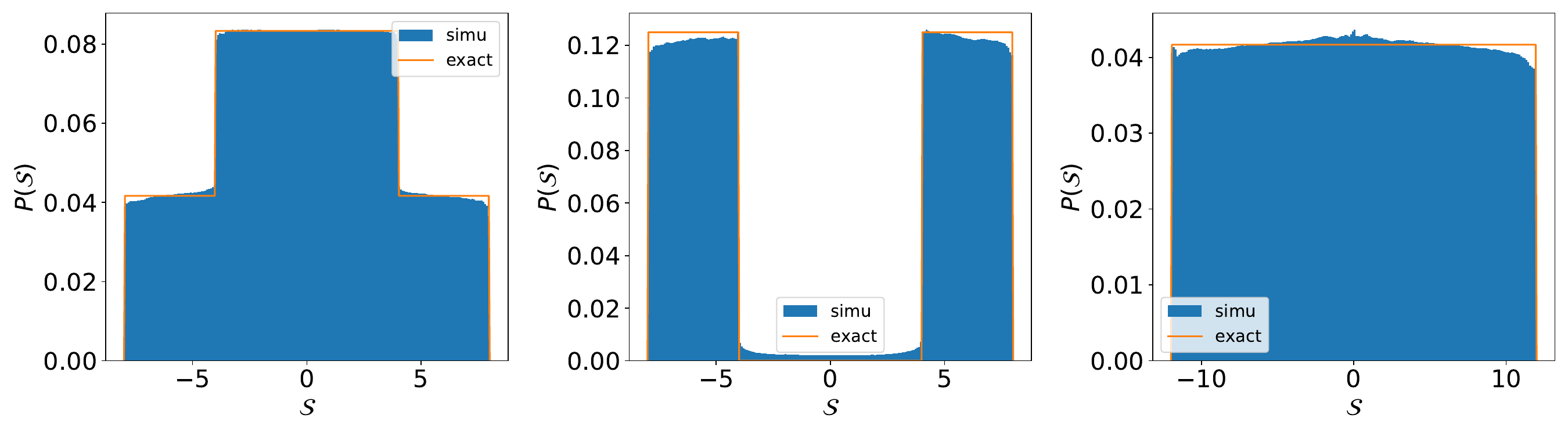} 
\caption{Vanishing switching rate $\lambda_x$. The probability density function $P(\mathcal{S})$ 
	as function of $\mathcal{S}$ for $u=1/2$,  $\alpha_x = 0$ and $\alpha_y = 1$. Left panel: $P(\mathcal{S})$ for  $v_x = 1$ and $v_y = 3$. Middle panel: $P(\mathcal{S})$ for  $v_x = 3$ and $v_y = 1$. Note that $P(\mathcal{S}) \equiv 0$ for $|\mathcal{S}| < \Omega $.
	Right panel: $P(\mathcal{S})$ for  $v_x = v_y = 3$. }
\label{fig:13} 
\end{figure} 

In Fig. \ref{fig:13} we plot $P(\mathcal{S})$ in eq. \eqref{pUalphax0} for $v_x = 1$ and $v_y = 3$ (left panel), $v_x = 3$ and $v_y = 1$ (middle panel) and  for $v_x = v_y$ (right panel). The PDF in the left panel is a sum of two rectangular functions: it has a flat terrace in the center, followed by two lower terraces on both sides, and drops to zero at $\mathcal{S} = \pm \Sigma$. The PDF presented in the middle panel consists of two rectangles separated by the interval $\mathcal{S} \in (-\Omega,\Omega)$ in which $P(\mathcal{S})$ is exactly equal to zero. Also in this regime we may expect that heavy tails of the PDF of the angular velocity of the SG may be absent. In the right panel we depict $P(\mathcal{S})$ in case of equal amplitudes of noises,  $v_x = v_y =3$.  In this case, the PDF is a single rectangular function in the interval $\mathcal{S} \in (-\Sigma,\Sigma)$ and equals zero outside of this interval. Note finally that some slight discrepancies between the analytical steady-state forms of the PDFs and the numerical simulations results are essentially due to the finite-time effects. In this particular case one has to take $t$ very large indeed in the numerical solutions of eqs. \eqref{LE} which are beyond reach.   

\begin{figure}[h!]
\centering
\includegraphics[width=150mm]{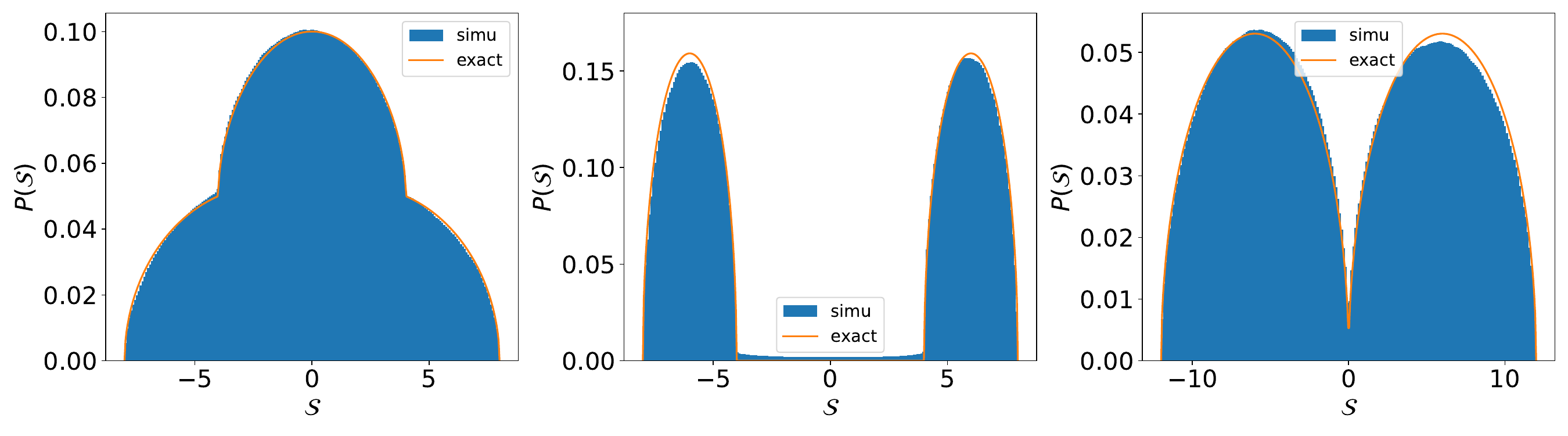} 
\caption{Vanishing switching rate $\lambda_x$. The probability density function $P(\mathcal{S})$ 
	as function of $\mathcal{S}$ for $u=1/2$, $\alpha_x = 0$ and $\alpha_y = 3/2$. Left panel: $P(\mathcal{S})$ for  $v_x = 1$ and $v_y = 3$. Middle panel: $P(\mathcal{S})$ for  $v_x = 3$ and $v_y = 1$. Note that $P(\mathcal{S}) \equiv 0$ for $|\mathcal{S}| < \Omega $. Right panel: $P(\mathcal{S})$ for  $v_x = v_y = 3$.}
\label{fig:14} 
\end{figure} 

Lastly, in Fig. \ref{fig:14} we plot $P(\mathcal{S})$ as function of $\mathcal{S}$ for $\alpha_y = 3/2$ - a particular instance of the case C. Here, for $v_x < v_y$ (left panel) the PDF is a piecewise continuous function, but its derivative is evidently discontinuous at points $\mathcal{S} = \pm \Omega$. In the middle panel we depict the PDF for the opposite case when $v_x > v_y$, where the PDF has the form of two bell-shaped functions separated by a finite interval in which the PDF equals zero. In such a situation one may also expect that the heavy tails of the PDF $P(W)$ of the angular velocity of the SG are absent.	
In the right panel,  for $v_x = v_y$, the two bell-shaped functions partially overlap forming a cusp-like minimum at $\mathcal{S} = 0$. 

\subsection{Vanishing amplitude $v_x$} 

\begin{figure}[H]
	\centering
\includegraphics[width=150mm]{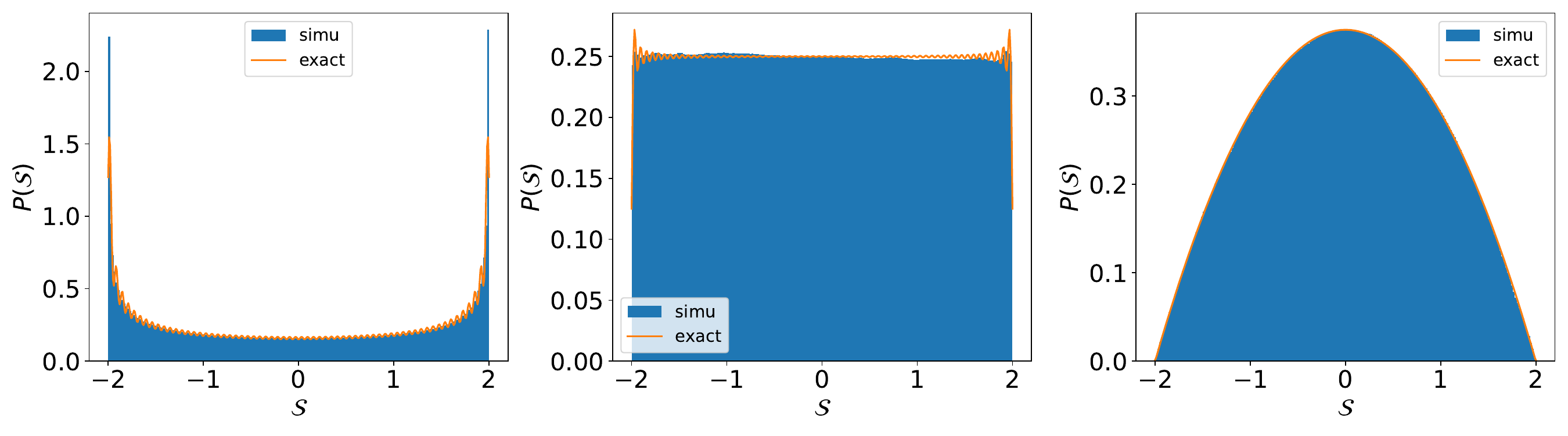} 		\caption{Vanishing amplitude $v_x$. The PDF $P(\mathcal{S})$ 
		as function of $\mathcal{S}$ for $u=1/2$, $v_x = 0$, $v_y = 1$  and $\alpha_x=1/2$. Left panel: $P(\mathcal{S})$ for $\alpha_y=1/2$.   Middle panel: $P(\mathcal{S})$ for  $\alpha_y=1$.  Right panel: $P(\mathcal{S})$ for  $\alpha_y=2$.}
	\label{fig:17} 
\end{figure}

We consider next the case when $v_x \equiv 0$ (i.e., $\zeta_x(t) \equiv 0$) and $v_y > 0$, such that the $x$-component can be considered as "passive" or cold, while the $y$-component is  "active" or hot, entraining the $x$-component into a stochastic evolution. In this case, the PDF in eq. \eqref{pU} becomes
\begin{align}
	\label{sancho2}
	P(\mathcal{S}) = \frac{(1-u) \Gamma(1/2+\alpha_y)}{\sqrt{\pi} \Gamma(\alpha_y) v_y^{2 \alpha_y -  1}}
	\begin{cases}
		\left(v_y^2 - (1 - u)^2 \mathcal{S}^2\right)^{\alpha_y -  1},  \,\,\,  \quad |\mathcal{S}| \leq v_y/(1 - u) \,,\\
		0, \qquad  \qquad \qquad \,\,\, \,\,\,\,\,\,\,\,\,\,\,\,\,\,\, \, \quad |\mathcal{S}| > v_y/(1 - u) \,, 
	\end{cases}
\end{align}
and has exactly the same functional form as the result in eq. \eqref{sancho}. Therefore, for $v_x = 0$ the PDF is a $U$-shaped function diverging at the edges of the support for $\alpha_y < 0$, is a constant for $\alpha_y = 1$ and $\mathcal{S} \in (-v_y/(1 - u),v_y/(1 - u))$, and
 is a bell-shaped function with a finite support for $\alpha_y > 1$. These three forms of the PDF are depicted in Fig.\ref{fig:17}.

\section{Conclusions}\label{conc}

To conclude, we considered the stochastic dynamics of a particle on a plane in the presence of a confining
parabolic potential, an analog of the widely used and experimentally relevant Brownian Gyrator
(BG) model. In contrast to the standard BG model with Gaussian  white-noises, we supposed that the time evolution of the particle's position is driven by two statistically independent \textit{dichotomous} noises  with arbitrary unequal amplitudes $v_x$ and $v_y$ and arbitrary unequal switching rates $\lambda_x$ and $\lambda_y$. In mathematical terms, the model --  which we call the Stochastic Gyrator (SG) -- is framed in terms of two linearly coupled (with the coupling parameter $u$, $|u| < 1$) generalized Ornstein-Uhlenbeck processes driven by dichotomous noises. 

We obtained analytical expressions for the position's
coordinates variances, cross-correlations, and mean specific angular momentum of the SG. This allowed
us to establish the conditions under which the mean specific  angular momentum is not equal to zero and  the spontaneous rotational motion of the SG
around the origin arises. 
Complementing these results, we presented a numerical analysis of
the mean specific angular velocity, showing several interesting properties absent in the BG model. In addition, our analysis revealed a remarkable discontinuous dependence of the mean specific angular velocity on the coupling parameter 
$u$, a feature that, to our knowledge, has not been reported previously. 

We also calculated analytically certain marginals of the joint position probability density of the SG, uncovering their remarkably rich (as compared to the standard BG model) behavior, that emerges in a seemingly
simple system of two coupled linear stochastic differential equations. In particular, we found
that depending on the noise parameters, these distributions undergo a series of noise-driven
transitions while approaching the steady-state, attaining in the limit $t \to \infty$ quite unusual shapes with 
multiple extrema, plateaus, and discontinuities. As well, we have shown that in several regimes the  stationary position probability 
density function is equal  to zero in an extended region 
around the origin suggesting that in the SG case the probability density function of the specific angular velocity, which exhibits power-law tails in the standard BG model, may possess all moments and consequently,  a heat engine with a more systematic performance can be indeed realized.  

In future work, we plan to extend this analysis by focusing on the joint position probability density function of the SG. A detailed understanding of this quantity will allow us to obtain the complete probability density functions of the specific angular momentum and specific angular velocity -- key observables that characterize the stochastic rotational dynamics of the SG driven by dichotomous noises. Such an analysis will provide a more comprehensive picture of how the interplay between noise switching and coupling  between the components governs the emergence of stochastic rotational behavior in such a non-equilibrium system.

\section*{Acknowledgements}

TH  wishes to thank Laboratoire de Physique Th\'eorique de la Mati\`{e}re Condens\'ee and its staff for a very warm hospitality during his $M1$ internship in June-July 2025.

\appendix
\section{A streamlined derivation of Eqs. (25) -- (26)}\label{A}

\vskip1cm According to (23), we have to obtain the stationary probability
density $P(s)$ of 
\begin{equation}
	s(t)=v\int_{0}^{t}\ e^{-\beta (t-\tau )}\zeta (\tau )d\tau ,\;\beta =(1-u),
	\label{xt}
\end{equation}%
where $\zeta (t)$ is the dichotomous random process, assuming values $\pm v$
with the rate $\pm \lambda $.

\medskip We will use the following general fact of the theory of Markov
processes.

\begin{claim}
	Let $\eta (t)$ be a Markov process (not necessarily dichotomous and possibly
	multicomponent) with the generator (infinitesimal operator) $L_{\eta }$
	i.e., its probability distribution $P(t,\eta )$ solves the Fokker-Planck
	equation 
	\begin{equation}
		\frac{\partial }{\partial t}P(t,\eta )=L_{\eta }P(t,\eta ),  \label{fpz}
	\end{equation}%
	Consider the differential equation%
	\begin{equation}
		\overset{.}{s}(t)=B(s(t),\eta (t)).  \label{xd}
	\end{equation}%
	Then the pair $(x(t),\eta (t))$ is also a Markov process and its joint
	probability density $P(t,s,\eta )$ solves the Fokker-Planck equation 
	\begin{equation}
		\frac{\partial }{\partial t}P=-\frac{\partial }{\partial s}(B(s,\eta )P%
		\mathcal{)+}L_{\eta }P.  \label{fpgen}
	\end{equation}
\end{claim}

\smallskip The proof of the claim is given at the Remark at the end of the
note.

\medskip Writing now (\ref{xt}) in the form (\ref{xd}), i.e., as $ \overset{.%
}{s}(t)=-\beta s(t)+\zeta (t)), $ we see that in our case (\ref{xt}) $%
\eta(t)-\zeta(t)$ and $B(s,\zeta )=-\beta s+\zeta $. Thus, taking into
account that for the dichotomous process

\begin{equation}
	(L_{\zeta }\mathcal{P})(t,\zeta )=-\lambda \mathcal{P}(t,+\zeta)+\lambda 
	\mathcal{P}(t,-\zeta),\; \zeta=\pm v,  \label{lze}
\end{equation}%
(see the Remark at the end of this note), we find that in this case (\ref%
{fpgen}) is%
\begin{equation}
	\frac{\partial }{\partial t}\mathcal{P}(t,s,\zeta)=\frac{\partial }{\partial
		s}((\beta s-\zeta )\mathcal{P}(t,s,\zeta))-\lambda \mathcal{P}%
	(t,s,\zeta)+\lambda \mathcal{P}(t,s,-\zeta), \; \zeta=\pm v.  \label{fpt}
\end{equation}%
Denote for convenience the stationary probability density of $(s(t),\zeta
(t))$ by $P_{\zeta }(s)=\mathcal{P}(\infty ,s,\zeta )$. It solves the
stationary version of (\ref{fpt})%
\begin{equation}
	\frac{\partial }{\partial s}((\beta s-\zeta )P_{\zeta }\mathcal{)}-\lambda
	P_{\zeta }+\lambda P_{-\zeta }=0,\;\zeta =\pm 1.  \label{fpst}
\end{equation}%
of (\ref{fpt}).

Set next $\mathcal{P=}P_{+}+P_{-},\;\mathcal{Q}=P_{+}-P$ and note that $%
\mathcal{P}=P$ is the stationary probability density of $s(t)\,\ $ of (\ref%
{xt}) that we are looking for.

Now we add and subtracts the equations in (\ref{fpst}) and obtain%
\begin{equation*}
	\frac{\partial }{\partial x}(\beta x\mathcal{P}-v\mathcal{Q)}=0,\;\frac{%
		\partial }{\partial x}(\beta x\mathcal{Q}-v\mathcal{P)}-2\lambda \mathcal{Q}%
	=0.
\end{equation*}%
Note now that (\ref{xt}) implies the bound: $|s(t)| \le v\beta^{-1}$  (cf.
(29)). It follows then that the probability density of (\ref{xt}), as well
as $\mathcal{P}$ and $\mathcal{Q}$ are zero is $|s|>v\beta^{-1}$. Thus, the
first equation above is equivalent to $\beta x\mathcal{P}-v\mathcal{Q=}0$
yielding $ \mathcal{Q=}\frac{\beta x}{v}\mathcal{P}$. Plugging this into the
second equation, we obtain%
\begin{equation}  \label{fin}
	(a\mathcal{P})^{\prime }-b\mathcal{P=}0,\;a(s)=(\beta
	s)^{2}-v^{2},\;b(s)=2\beta \lambda s.
\end{equation}%
Solving this ODE, using the normalization condition%
\begin{equation*}
	\int \mathcal{P(}x\mathcal{)}dx=1,
\end{equation*}%
and setting $\beta=1-u,\; \lambda=\lambda_x, \; v=v_x\sqrt{i}$, we obtain formulas (25) -- (26) of the manuscript.

\medskip \textit{Remark}. Here we prove the Claim above. We start with the
another definition of the infinitesimal operator $L_{\eta }$ in (\ref{fpz}).
Consider $\ \mathbb{E}_{\eta _{0}}\{f(\eta (t))\},\;t\geq 0,$ where $\mathbb{%
	E}_{\eta _{0}}\{...\}$ is the expectation over the realizations of $\eta
(t),\;\eta (0)=\eta _{0}.$ Then the operator $L_{\eta }^{\ast }$, that is
conjugate to $L_{\eta }$ in the r.h.s. of (\ref{fpz}), is defined by 
\begin{equation}
	\ \mathbb{E}_{\eta _{0}}\{f(\eta (\delta ))\}=\ f(\eta _{0})\}+\delta
	(L_{\eta }^{\ast }f)(\eta _{0})+o(\delta ),\;\delta \rightarrow 0.
	\label{lcon}
\end{equation}%
In particular, taking into account that during the small interval $(0,\delta
)$ the dichotomous process stays unchanged with probability $1-\lambda
\delta $ and changes its state to another one with probability $\lambda
\delta $ and using (\ref{lcon}), we obtain (\ref{fpz}).

Consider now the process $(s(t),\eta (t))$ defined by $\eta (t)$ and (\ref%
{xd}). Note first that according to (\ref{xd}) \ 
\begin{equation}
	s(\delta )=s_{0}+B(s_{0},\eta _{0})\delta +o(\delta ).  \label{dex}
\end{equation}%
Now, writing

\begin{align*}
	& \mathbb{E}_{s_{0},\eta _{0}}\{f(s(\delta ),\eta (\delta ))\}=f(s_{0},\eta
	_{0})+\mathbb{E}_{s_{0},\eta _{0}}\{f(s(\delta ),\eta (\delta
	))-f(s_{0},\eta (\delta ))\} \\
	& \hspace{7cm}+\mathbf{E}_{s_{0},\eta _{0}}\{f(s_{0},\eta (\delta
	))-f(s_{0},\eta _{0})\},
\end{align*}%
for some function $f$ and using (\ref{dex}) in the first term on the right
and (\ref{fpz}) in the second term on the right, we obtain%
\begin{align*}
	& \mathbb{E}_{s_{0},\eta _{0}}\{f(s(\delta ),\eta (\delta ))\}-f(s_{0},\eta
	_{0}) \\
	& =B(s_{0},\eta _{0})\frac{\partial }{\partial s_{0}}f(s_{0},\eta
	_{0})+(L_{\eta }^{\ast }f)(s_{0},\eta _{0})+o(\delta )=(L^{\ast
	}f)(s_{0},\eta _{0})++o(\delta ),
\end{align*}%
where%
\begin{equation*}
	L^{\ast }=(B(s,\eta )\frac{\partial }{\partial s}+L_{\eta }^{\ast }.
\end{equation*}%
The operator in the r.h.s. of (\ref{fpgen}) is obviously conjugate to $%
L^{\ast }$ above.

\section{Mixed case: $\zeta_x(t)$ is a Gaussian white-noise while $\zeta_y(t)$ is a dichotomous noise}
\label{GG}

Consider finally a mixed case in which either of the noises, say $\zeta_x(t)$, is a Gaussian white-noise, while the second one, $\zeta_y(t)$, is a dichotomous noise. We note that this case can be experimentally realized. In fact, in the representation of the standard BG model in terms of an equivalent electric circuit two thermal bathes were introduced via independent noises generated by a computer acting on both resistances. It does not look like a big problem to incorporate in this way the noises with arbitrary prescribed properties. 

The PDF $P(\mathcal{S})$ can be derived straightforwardly from our eq. \eqref{pU} by first exponentiating the contribution stemming out of averaging over $\zeta_x(t)$ and then turning to the double limit $\lambda_x \to \infty$ and $v_x \to \infty$ with the ratio $D_x = v_x^2/2 \lambda_x$ in eq. \eqref{diffD} kept fixed. In doing so, we find     
\begin{align}
	\begin{split}
		\label{pUU}
		P(\mathcal{S}) &= \frac{\Gamma(1/2 + \alpha_y)}{\pi }  \left(\frac{v_y}{2 (1-u)}\right)^{1/2 - \alpha_y}  \\
		& \times \int^{\infty}_0 d\omega \,  \exp\left( - \frac{D_x \, \omega^2}{2 (1- u)}\right) \, \dfrac{J_{\alpha_y - 1/2}\left(\dfrac{v_y}{1 - u} \omega\right)}{\omega^{\alpha_y - 1/2}} \, \cos(\omega \mathcal{S})  \,,
	\end{split}
\end{align}
Note that since $v_x = \infty$, also $\Sigma = \infty$, i.e., the support of  $P(\mathcal{S})$ is the entire real line.

We are unable to perform the integral in eq.\eqref{pUU} exactly for arbitrary values of $\alpha_y$. Therefore, we will consider instead two particular cases: $\alpha_y = 1/2$ which is analysed via  numerical simulations, and the case $\alpha_y =1$, in which the integral in eq. \eqref{pUU} can be performed exactly. These two situations will suffice  to show that in this mixed Gaussian noise / dichotomous noise case $P(\mathcal{S})$ does not attain a simple Gaussian form in eq. \eqref{Gauss} but still exhibits  a rather non-trivial behavior. 
Then, 
we will derive, for arbitrary $\alpha_y$,  the short- and large-$\mathcal{S}$ asymptotic series expansions of $P(\mathcal{S})$.

\begin{figure}[h!]
	\centering	
	\includegraphics[width=150mm]{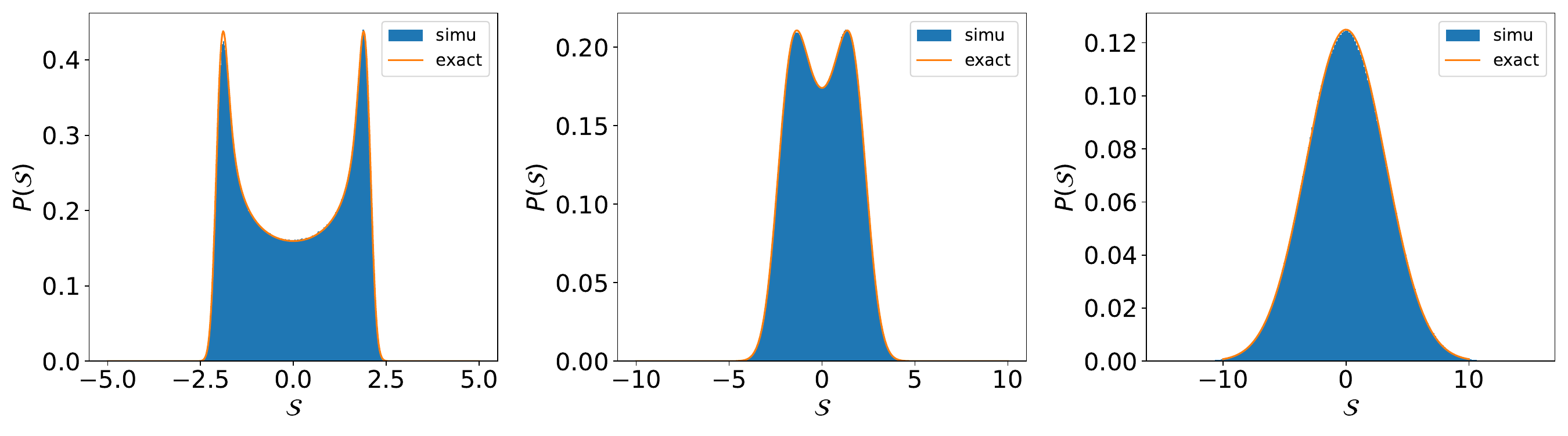}	
	\caption{Mixed case: Gaussian white-noise and dichotomous noise acting together. The probability density function $P(\mathcal{S})$
		as function of $\mathcal{S}$ for $u=1/2$ and $\alpha_y = 1/2$.  Left panel: $P(\mathcal{S})$ for  $D_x = 0.01$ and $v_y = 1$. Middle panel: $P(\mathcal{S})$ for  $D_x = 0.25$ and $v_y = 1$ Right panel: $P(\mathcal{S})$ for  $D_x = 4$ and $v_y = 1$.}
	\label{fig:x}
\end{figure}

For $\alpha_y = 1/2$ and $\zeta_x(t)$ being the Gaussian white-noise, (which can be also considered as an extreme example of the case C),  the PDF in eq. \eqref{pUU} can be conveniently represented, taking advantage of  the standard integral representation of the Bessel function $J_0(\ldots)$ \cite{gradshteyn2007}, as
\begin{equation}
	\begin{split}
		\label{asy5b}
		P(\mathcal{S}) &=  \sqrt{\frac{(1-u)}{2 \pi^3 D_x}}\int_0^{2\pi} d\theta    \exp\left(-\frac{(1-u)}{2D_x}\left(\frac{\sin(\theta ) v_y}{1-u}+S\right)^2\right) \,.
	\end{split}
\end{equation}
Expanding the exponential function into the Taylor series in powers of $\mathcal{S}$ and integrating, we find that $P(\mathcal{S})$   has a parabolic form in the neighborhood  of $\mathcal{S} =0$ :
\begin{align}
	\begin{split}
		\label{as11}
		P(\mathcal{S}) &= \sqrt{\frac{2 (1-u)  D_x^2}{\pi}} \exp\left(- \frac{v_y^2}{4 (1- u) D_x}\right) I_0\left(\frac{v_y^2}{4 (1-u)D_x}\right) \Bigg[1 - \Bigg(\frac{(1- u)}{2D_x} +  \\&+ \frac{v_y^2}{4 D_x^2} \left(I_1\left(\dfrac{v_y^2}{4 (1-u)D_x}\right)/I_0\left(\dfrac{v_y^2}{4 (1-u)D_x}\right) - 1
		\right)\Bigg) \mathcal{S}^2 + O\left(\mathcal{S}^4\right)
		\Bigg]	\,.	
	\end{split}
\end{align}
Further on, some standard analysis shows that in the large-$|\mathcal{S}|$ limit the expression \eqref{asy5b} attains the following asymptotic form : 
\begin{align}
	\label{as8}
	P(\mathcal{S}) \simeq \frac{1}{2 \pi} \sqrt{\frac{(1-u)}{v_y |\mathcal{S}|}} \exp\left(- \frac{(1-u) \mathcal{S}^2}{2 D_x} + \frac{v_y |\mathcal{S}|}{D_x} - \frac{v_y^2}{2 (1 - u) D_x}\right) \,,
\end{align}
where the symbol $\simeq$ signifies that eq. \eqref{as8} presents only the leading-order behavior and all the subdominant terms are omitted. Here, the following remark is in order: 
one may expect, on intuitive grounds, that the Gaussian white-noise with its unbounded amplitude will completely dominate the form of $P(\mathcal{S})$ for large values of $\mathcal{S}$: indeed, the support of $P(\mathcal{S})$ in eq. \eqref{asy5b} (and, in general, in eq. \eqref{pUU}) is the entire real line because $v_x = \infty$, and also in the above asymptotic expression \eqref{as8} the decay of $P(\mathcal{S})$ with $\mathcal{S}$
is governed by the Gaussian function $\exp(- (1- u)\mathcal{S}^2/2 D_x)$, in which only the "diffusion coefficient" $D_x$ determines the characteristic scale. 	However, the subdominant, exponentially growing term $\exp(v_y |\mathcal{S}|/D_x)$ is also non-negligible and moreover, the amplitude of the large-$|\mathcal{S}|$ decay is dominated by the value of the amplitude  $v_y$ of the dichotomous noise. Quite counter-intuitively, both the Gaussian and the dichotomous noises also control the shape of $P(\mathcal{S})$ for small values of $\mathcal{S}$. In fact,  
the coefficient in front of $\mathcal{S}^2$ in eq. \eqref{as11}
can be positive or negative, meaning that  $\mathcal{S} = 0$ can be a maximum or a minimum, 
depending on the relation between the values of $v_y$ and $D_x$. This signifies that the dynamical transition which the system undergoes in the course of time is dominated by both noises. In Fig. \ref{fig:x} we depict the steady-state PDF in eq. \eqref{asy5b} (evaluated in Scipy\cite{Virtanen2020}) for three values of the diffusion coefficient $D_x$ and fixed amplitude $v_y = 1$ of the dichotomous noise. 

\begin{figure}[h!]
	\centering
	\includegraphics[width=150mm]{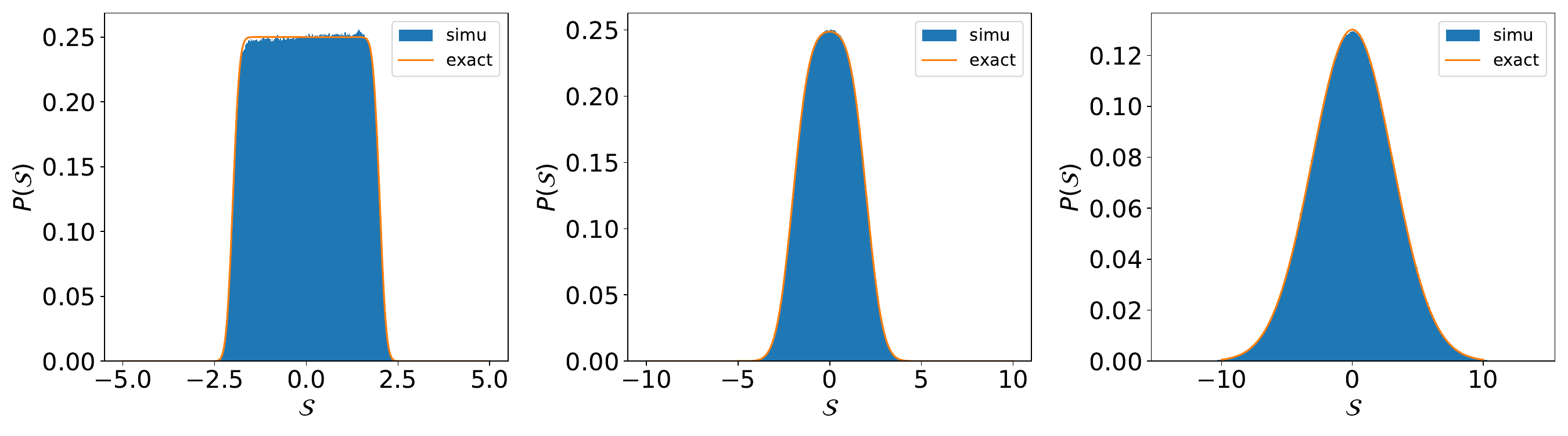}			\caption{Mixed case: Gaussian white-noise and dichotomous noise acting together.  The probability density function $P(\mathcal{S})$
		as function of $\mathcal{S}$ for $u=1/2$ and $\alpha_y = 1$. Left panel: $P(\mathcal{S})$ for  $D_x = 0.01$ and $v_y = 1$. Middle panel: $P(\mathcal{S})$ for  $D_x = 0.25$ and $v_y = 1$ Right panel: $P(\mathcal{S})$ for  $D_x = 4$ and $v_y = 1$.}
	\label{fig:xx}
\end{figure}

Consider next the case $\alpha_y = 1$ (which can be seen as some extreme limiting situation for the case E), in which the integral in eq. \eqref{pUU} can be performed exactly. 
We find that here $P(\mathcal{S})$ is given explicitly by
\begin{align}
	\label{small}
	P(\mathcal{S}) = \frac{1 - u}{4 v_y} \left[{\rm erf}\left(\sqrt{\frac{1-u}{2 D_x}} \left( \frac{v_y}{1-u} -  \mathcal{S} \right)\right) + {\rm erf}\left(\sqrt{\frac{1-u}{2 D_x}} \left( \frac{v_y}{1-u} +  \mathcal{S} \right)\right)  \right] \,,
\end{align}
where ${\rm erf}(\ldots)$ is the error function \cite{gradshteyn2007}. The PDF in eq. \eqref{small} shows the following asymptotic behavior :
\begin{align}
	\begin{split}
		\label{asy1}
		P(\mathcal{S}) &= \frac{1-u}{2 v_y} {\rm erf}\left(\frac{v_y}{\sqrt{2 (1-u) D_x}}\right) - \sqrt{\frac{(1 - u)^3}{8 \pi D_x^3}} \exp\left(- \frac{v_y^2}{2 (1-u) D_x}\right) \mathcal{S}^2 + O\left(\mathcal{S}^4\right) \,, 
	\end{split}
\end{align}
for $|\mathcal{S}| \ll v_y/(1-u)$, while in the opposite limit $|\mathcal{S}| \to \infty$ one has
\begin{align}
	\label{ju}
	P(\mathcal{S}) &\simeq \sqrt{\frac{(1-u) D_x}{2 \pi v_y^2}}  \exp\left(- \frac{(1-u)}{2 D_x} \mathcal{S}^2 - \frac{v_y^2}{2 (1-u) D_x}\right) \dfrac{\sinh\left(\dfrac{v_y S}{D_x}\right)}{\mathcal{S}}  \,.
\end{align}
Therefore, $P(\mathcal{S})$ also has a parabolic form for small $\mathcal{S}$. Note that the coefficient in front of $\mathcal{S}^2$ is always positive such that $\mathcal{S} = 0$ is always the maximum of the PDF. However, the coefficient in front of $\mathcal{S}^2$ in eq. \eqref{asy1} is
a strongly decreasing function of the amplitude $v_y$ of the dichotomous noise, which implies that for large enough $v_y$ the PDF may have a plateau in its central part. Emergence of such a plateau is a generic feature of the situation in which either of $\alpha$-s is equal to unity, as we have evidenced above. Further on,  we observe that 
in the limit $|\mathcal{S}| \to \infty$ the behavior appears to be very similar to the one observed above for $\alpha_y = 1/2$:  
the dominant rate-controlling factor which governs the
decay of $P(\mathcal{S})$ is the Gaussian function $\exp(-(1-u)  \mathcal{S}^2/2 D_x)$.  However, 
the dichotomous noise contribution is still important in this limit - it defines the amplitude of the decay and also the subdominant, growing with $\mathcal{S}$ term - $\sinh(v_y \mathcal{S}/D_x)/\mathcal{S}$,  which are important for any finite $\mathcal{S}$. Consequently, both the short- and large-$\mathcal{S}$ behaviors of the PDF result from quite an intricate interplay of both noises. We depict the forms of the PDF $P(\mathcal{S})$ in Fig. \ref{fig:xx} for three different values of $D_x$ and fixed value of the amplitude $v_y = 1$ of the dichotomous noise. 

We turn next to the case of arbitrary $\alpha_y$ aiming to determine the asymptotic behavior of $P(\mathcal{S})$.
This can be readily done by noticing that a convenient series representation of the product of a cosine and of the Bessel function divided by a power of $\omega$ can be obtained directly from eq. \eqref{ser}, upon setting either $\mu$ or $\nu$ equal to $-1/2$. This gives, for $|\mathcal{S}| \leq v_y/(1-u)$, the  following expansion
\begin{align}
	\begin{split}
		\label{exp1}
		\dfrac{J_{\alpha_y - 1/2}\left(\dfrac{v_y }{1- u} \omega\right)}{\left(\dfrac{v_y}{2 (1 - u)} \omega\right)^{\alpha_y - 1/2}}	\cos(\omega \mathcal{S}) &= \sum_{k=0}^{\infty} \dfrac{(-1)^k \left(\dfrac{v_y}{2 (1-u)} \omega\right)^{2 k} }{k! \Gamma(k+ \alpha_y + 1/2)} \\
		&\times \,_2F_1\left(-k, -k - \alpha_y +1/2; 1/2; \frac{(1-u)^2 \mathcal{S}^2}{v_y^2}\right) \,.
	\end{split}
\end{align}
Inserting the expansion \eqref{exp1} into eq. \eqref{pUU} and integrating, we get
\begin{align}
	\begin{split}
		\label{exp3}
		P(\mathcal{S}) &= \frac{\Gamma(1/2+\alpha_y)}{\pi} \sqrt{\frac{1-u}{2 D_x}} \sum_{k=0}^{\infty} \dfrac{(-1)^k \Gamma(k+1/2) }{k! \Gamma(k+ \alpha_y + 1/2)} \left(\frac{v_y^2}{2 (1-u) D_x}\right)^{k} \\
		&\times \,_2F_1\left(-k, -k - \alpha_y +1/2; 1/2; \frac{(1-u)^2 \mathcal{S}^2}{v_y^2}\right) \,,
	\end{split}
\end{align}
from which we readily get the short-$\mathcal{S}$ expansion of the form
\begin{align}
	\begin{split}
		P(\mathcal{S}) &= \sqrt{\frac{1-u}{2 \pi D_x}} \,_1F_1\left(1/2; 1/2 + \alpha_y; - \frac{v_y^2}{2 (1-u)D_x}\right) \\&- \sqrt{\frac{(1-u)^3}{8 \pi D_x^3}} \Bigg[\left(1 - \frac{v_y^2}{(1-u) D_x}\right) \,_1F_1\left(1/2; 1/2 + \alpha_y; - \frac{v_y^2}{2 (1-u)D_x}\right)\\
		&+ \frac{2 \alpha_y }{(1 + 2 \alpha_y)} \frac{v_y^2}{(1 - u)  D_x} \,_1F_1\left(1/2; 3/2 + \alpha_y; - \frac{v_y^2}{2 (1-u)D_x}\right) \Bigg].
	\end{split}
\end{align}	
The expression \eqref{small} follows from the above general short-$\mathcal{S}$ expansion if we set $\alpha_y = 1$.

Further on, a convenient large-$\mathcal{S}$ representation of $P(\mathcal{S})$ can be obtained as follows: Using the standard Taylor series expansion of the Bessel function, inserting it into eq. \eqref{pUU} and integrating, we get
\begin{align}
	P(\mathcal{S}) = \Gamma(\alpha_y +1/2) \sqrt{\frac{1-u}{2 \pi D_x}} \exp\left(- \frac{(1-u) \mathcal{S}^2}{2 D_x}\right) \sum_{k=0}^{\infty} \dfrac{\left(\dfrac{v_y^2}{8 (1-u) D_x}\right)^k}{k! \Gamma(k + \alpha_y + 1/2)} {\rm H}_{2 k}\left(\sqrt{\frac{1 - u}{2 D_x}} \mathcal{S}\right) \,,
\end{align}
where ${\rm H}_{2 k}(\ldots)$ are the Hermite polynomials \cite{gradshteyn2007}. Inserting into the latter expansion the  formal definition of the Hermite polynomials as a finite sum, $H_{2k}(x) = \sum_{p}^k c_{p,k} x^{2 p}$, interchanging then the summation over $k$ and $p$ and eventually retaining only the leading in the limit $\mathcal{S} \to \infty$ terms, we find
\begin{align}
	\begin{split}
		P(\mathcal{S}) &= \Gamma(\alpha_y +1/2) \sqrt{\frac{1-u}{2 \pi D_x}} \left(\frac{v_y \mathcal{S}}{2 D_x}\right)^{1/2-\alpha_y} \exp\left(- \frac{(1-u) \mathcal{S}^2}{2 D_x}\right) 	\\
		&\times \sum_{p=0}^{\infty} \frac{(-1)^p}{p!} \left(\frac{v_y^2}{2 (1-u) D_x}\right)^p I_{2 p + \alpha_y - 1/2}\left(\frac{v_y \mathcal{S}}{D_x}\right) \,,
	\end{split}
\end{align}
where $I_{\nu}(\ldots)$ is the modified Bessel function \cite{gradshteyn2007}. Finally, noticing  that for large values of the argument the modified Bessel function all follow
\begin{align}
	I_{2 p + \alpha_y - 1/2}\left(z\right) \simeq \dfrac{e^{|z|}}{\sqrt{2 \pi z}} \,,
\end{align}
we eventually obtain the general asymptotic result of the form
\begin{align}
	\begin{split}
		P(\mathcal{S}) \simeq \frac{\Gamma(\alpha_y +1/2)}{2 \pi} \sqrt{\frac{1-u}{2 \pi D_x}} \left(\frac{2 D_x}{v_y |\mathcal{S}|}\right)^{\alpha_y} \exp\left(- \frac{(1-u) \mathcal{S}^2}{2 D_x} + \frac{v_y |\mathcal{S}|}{D_x} - \frac{v_y^2}{2 (1-u) D_x}\right) 	 \,.
	\end{split}
\end{align}
Setting $\alpha_y =1$ we recover from the latter expression the result in eq. \eqref{ju}, while setting $\alpha_y = 1/2$ we find the asymptotic form in eq. \eqref{as8}.

\section*{References}


\providecommand{\newblock}{}

\end{document}